\documentclass[twocolumn]{aastex631}
\accepted{for publication in The Astronomical Journal on August 18th, 2023}

\usepackage{textgreek}

\shorttitle{Clustering properties of intermediate and high-mass YSOs}
\shortauthors{Vioque et al.}
\graphicspath{{./}{figures/}}

\begin{document}

\title{Clustering properties of intermediate and high-mass Young Stellar Objects\footnote{Table~\ref{Table} will be made available online at the CDS in the VizieR archive service.}}

\correspondingauthor{Miguel Vioque}
\email{miguel.vioque@alma.cl}

\author[0000-0002-4147-3846]{Miguel Vioque}
\affiliation{Joint ALMA Observatory, Alonso de Córdova 3107, Vitacura, Santiago 763-0355, Chile}
\affiliation{National Radio Astronomy Observatory, 520 Edgemont Road, Charlottesville, VA 22903, USA}

\author[0009-0002-2978-8383]{Manuel Cavieres}
\affiliation{Instituto de Astrofísica, Pontificia Universidad Católica de Chile, Av. Vicuña Mackenna 4860, 7820436 Macul, Santiago, Chile}

\author[0000-0001-9933-1229]{Michelangelo Pantaleoni González}
\affiliation{Centro de Astrobiología (CSIC-INTA), Departamento de Astrofísica, ESA-ESAC Campus, E-28691 Madrid, Spain}

\author[0000-0003-3133-3580]{Álvaro Ribas}
\affiliation{Institute of Astronomy, University of Cambridge, Madingley Road, Cambridge, CB3 0HA, UK}

\author[0000-0001-7703-3992]{René D. Oudmaijer}
\affiliation{ School of Physics and Astronomy, Sir William Henry Bragg Building, University of Leeds, Leeds LS2 9JT, UK}

\author[0000-0002-0233-5328]{Ignacio Mendigutía}
\affiliation{Centro de Astrobiología (CSIC-INTA), Departamento de Astrofísica, ESA-ESAC Campus, E-28691 Madrid, Spain}

\author[0000-0002-0885-1198]{Lena Kilian}
\affiliation{School of Earth and Environment, University of Leeds, Woodhouse Lane, LS2 9JT, Leeds, UK}

\author[0000-0001-7668-8022]{Héctor Cánovas}
\affiliation{Telespazio UK for the European Space Agency (ESA), European Space Astronomy Centre (ESAC), Camino Bajo del Castillo s/n, E-28692 Villanueva de la Cañada, Madrid, Spain}

\author[0000-0002-0631-7514]{Michael A. Kuhn}
\affiliation{Centre for Astrophysics Research, University of Hertfordshire, College Lane, Hatfield, AL10 9AB, UK}






\begin{abstract}

We have selected 337 intermediate and high-mass YSOs ($1.5$ to $20$ M$_{\odot}$) well-characterised with spectroscopy. By means of the clustering algorithm HDBSCAN, we study their clustering and association properties in the Gaia DR3 catalogue as a function of stellar mass. We find that the lower mass YSOs ($1.5-4$ M$_{\odot}$) have clustering rates of $55-60\%$ in Gaia astrometric space, a percentage similar to the one found in the T Tauri regime. However, intermediate-mass YSOs in the range $4-10$ M$_{\odot}$ show a decreasing clustering rate with stellar mass, down to $27\%$. We find tentative evidence suggesting that massive YSOs ($>10$ M$_{\odot}$) often appear –yet not always– clustered. We put forward the idea that most massive YSOs form via a mechanism that demands many low-mass stars around them. However, intermediate-mass YSOs form in a classical core-collapse T Tauri way, yet they do not appear often in the clusters around massive YSOs. We also find that intermediate and high-mass YSOs become less clustered with decreasing disk emission and accretion rate. This points towards an evolution with time. For those sources that appear clustered, no major correlation is found between their stellar properties and the cluster sizes, number of cluster members, cluster densities, or distance to cluster centres. In doing this analysis, we report the identification of 55 new clusters. We present tabulated all the derived cluster parameters for the considered intermediate and high-mass YSOs.

\end{abstract}




\section{Introduction}\label{Sec_intro}

A significant fraction of the young stellar object (YSO) population in the Galaxy is spatially clustered to some degree (e.g. \citealp{2014ApJ...787..107K, 2018A&A...620A.172Z}) and appears associated with star forming regions and molecular clouds (\citealp{2000AJ....120.3139C,2003ARA&A..41...57L,2019MNRAS.483.4707G, 2019ARA&A..57..227K, 2021A&A...651L..10K}). This is expected from a theory of turbulent collapse driving the hierarchical evolution of molecular clouds (e.g. \citealp{2016ApJ...817....4F,2019MNRAS.486..283B, 2021MNRAS.506.3239G, 2022MNRAS.512..216G}). Often, YSOs can be found in co-moving groups of stars (e.g. \citealp{2022A&A...664A.175P}). Some of these groups can be fitted with precision by a single isochrone, thus hinting at a common origin (e.g. \citealp{2022A&A...661A.118C}). Nevertheless, evidence points against a simple linear star formation history in many of these groups, which are often gravitationally unbound and hierarchically distributed (\citealp{2014ApJ...787..107K,2019ApJ...870...32K,2019A&A...626A..35R, 2022MNRAS.515..167G,2022A&A...664A..66M, 2022MNRAS.516.5258S}).

Despite this clustering nature, all-sky systematic surveys have reported many YSOs in relative isolation. For the low-mass regime (M\textless{}$1.5$ M$_{\odot}$),  $\sim40-50\%$ of YSOs have been found relatively isolated (\citealp{2009ApJS..184...18G,2009ApJ...694..367S,2012ApJ...748...64G,2020AJ....160...68W,2021ApJS..254...33K}). Even within massive star forming regions, $\sim20\%$ of the YSOs appear unclustered (\citealp{2015ApJ...802...60K}). This number is consistent with theoretical expectations (e.g. \citealp{2012MNRAS.426.3008K, 2020MNRAS.494..624K}). These isolated YSOs are often the result of dispersed short-lived clusters, ejection from nearby clusters, isolated star formation, or a combination of all the previous phenomena.


A large scale analysis of the clustering properties of intermediate and high-mass YSOs (M\textgreater{}$1.5$ M$_{\odot}$) is still missing. At this mass regime, the isolated nature of some forming stars is of particular interest. These sources evolve much faster than their lower mass counterparts (e.g. \citealp{2012MNRAS.427..127B}), and thus there is significantly less time for them to be scattered in the field. In addition, they require larger reservoirs of material for their formation, which suggests a stronger connection with clustered environments. Indeed, more massive YSOs tend to appear more clustered (e.g., \citealp{1995PhDT.........1H, 1999A&A...342..515T}), and larger clusters have more massive YSOs (e.g. \citealp{2012ApJ...745..131K, 2021AJ....161..257K}). However, many intermediate and high-mass YSOs have been found relatively isolated, outside star forming regions, and not belonging to any stellar overdensity (\citealp{2008ApJ...678.1070S, 2012A&A...542A..49B,2023ApJ...942....7F,2023AJ....165....3K, 2022ApJ...939..120L}). Studies of OB stars also point in the direction that not all massive stars form in clusters (\citealp{2005A&A...437..247D,2020MNRAS.495..663W, 2022arXiv220310007W}). This puts important constrains to massive star formation theories, particularly in the case of the more massive stars (M\textgreater{}$10$ M$_{\odot}$). Relatively isolated massive YSOs are often indicative that a monolithic core collapse has happened (e.g. \citealp{2003ApJ...585..850M}), whereas the absence of a significant population of isolated massive YSOs would be supportive of a competitive accretion scenario dominating massive star formation (e.g. \citealp{2010ApJ...709...27W}).


In addition, environmental and clustering properties can significantly affect YSOs, particularly when in the proximity of massive stars (\citealp{2023EPJP..138..675L}). High density environments can affect their protoplanetary disks (e.g. \citealp{2021MNRAS.501.4317P, 2022MNRAS.tmp.1811C, 2022A&A...664A..66M, 2022EPJP..137.1132W}), stellar accretion (e.g. \citealp{2020MNRAS.497L..40W}), and planet forming mechanisms (e.g. \citealp{2020ApJ...905L..18K, 2020Natur.586..528W,2021ApJ...910L..19C, 2021ApJ...911L..16L}). In addition, larger stellar densities increase the chances of flybys, which can greatly affect star and planet formation (see \citealp{2023EPJP..138...11C}).

The advent of the Gaia mission (\citealp{2016A&A...595A...1G}) has brought a revolution in the study of stellar clusters and associations (e.g. \citealp{2020A&A...640A...1C}). In this paper, we make use of its Data Release 3 astrometry (\citealp{2021A&A...649A...2L}) and photometry (\citealp{2021A&A...649A...3R}), in combination with a new large sample of well-characterized sources (\citealp{2022ApJ...930...39V}), to present the first systematic study of the clustering properties of intermediate and high-mass YSOs. In Section~\ref{S_sample}, we present the sample of 337 intermediate and high-mass YSOs we consider in this study. In Section~\ref{S_HDBSCAN}, we describe how we have identified clusters and associations of Gaia DR3 sources containing these YSOs using the machine learning HDBSCAN algorithm. We analyse these clusters and the isolated YSOs in Sect.~\ref{S_Analysis}, including a comparison with published catalogues of open clusters. We discuss our findings in Sect.~\ref{S_Discussion}, and we conclude in Section~\ref{S_Conclusion}.

\begin{figure*}[t!]
\centering\includegraphics[width=\textwidth]{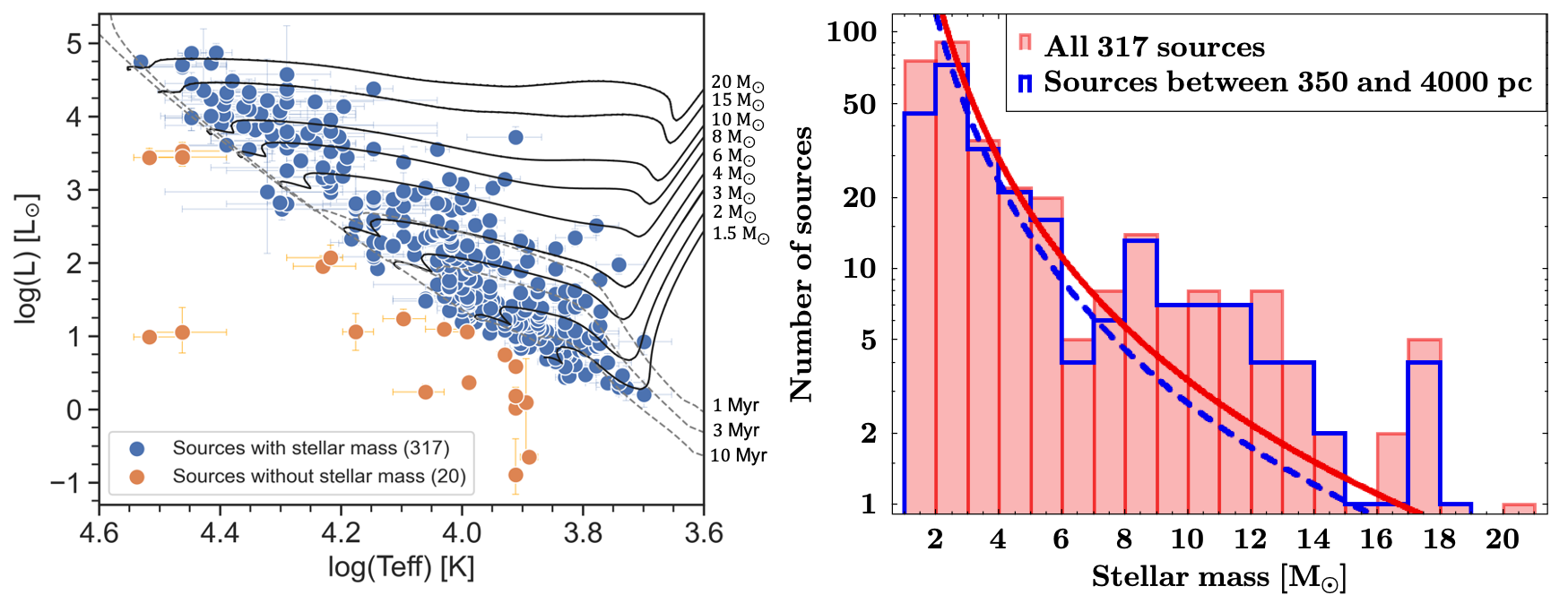}
\caption{\textit{Left:} Hertzsprung-Russell diagram of the considered sample of 337 intermediate and high-mas YSOs. PARSEC 1.2S PMS tracks and isochrones corresponding to $1.5$, $2$, $3$, $4$, $6$, $8$, $10$, $15$, $20$ M$_{\odot}$ and $1$, $3$, and $10$ Myr are presented (\citealp{2012MNRAS.427..127B} and \citealp{2017ApJ...835...77M}). \textit{Right:} Histogram of the number of YSOs considered in this work as a function of stellar mass. The lines indicate a \citet{1955ApJ...121..161S} IMF adjusted to the 2-3 M$_{\odot}$ bin.}\label{Plot: The sample}
\end{figure*} 


\section{Sample}\label{S_sample}



We compiled 337 well-characterised intermediate and high-mass YSOs (M\textgreater{}$1.5$ M$_{\odot}$) with derived stellar parameters from spectra and Gaia DR3 5-parameter astrometrical detections (\textalpha{}, \textdelta{}, $\varpi$, \textmu{}\textsubscript{\textalpha{}}\textsubscript{*}, \textmu{}\textsubscript{\textdelta{}}; i.e., right ascension, declination, parallax, proper motion right ascension, and proper motion declination). We selected 222 of these sources from the catalogues of historically considered and well-studied intermediate and high-mass YSOs of \citet{2018A&A...620A.128V} and \citet{2021A&A...650A.182G}. To these, we added 115 of the newly identified intermediate and high-mass YSOs of \citet{2022ApJ...930...39V}. The latter were selected from the \citet{2020A&A...638A..21V} catalogue, and hence are not biased towards any preferred location in the Galaxy (although \citealp{2020A&A...638A..21V} catalogue is limited to the Galactic plane, $-5<b<5$). 


We obtained two sets of distances from Gaia DR3 parallaxes for this sample of 337 sources, and compare them against each other. The first set concerns the geometric distances of \citet{2021AJ....161..147B}. For the second set, we used a Bayesian approach with a prior for massive stars (see \citealp{2021MNRAS.504.2968P}). Both distance sets are in great agreement, only showing significant differences for five (2\%) of the sources. Indeed, 98\% of the sample has $\varpi/\sigma(\varpi)>3$ and 84\% of the sample has \textit{ruwe} $<2$ (see \citealp{2021A&A...649A...2L} for Gaia goodness-of-fit parameters). We conclude that our distances are accurate within error bars and independent of the underlying distance prior. Because of this accuracy, from now on we only consider the \citet{2021AJ....161..147B} distances. The main reason for this is that the other distance set assumes a massive star (\textgreater{}$10$ M$_{\odot}$) prior, and all the five discordant sources are known to be below this threshold.


For accuracy, only effective temperatures derived from spectra are considered (in order of precision, we use those listed in: \citealp{2020MNRAS.493..234W, 2015MNRAS.453..976F, 2022ApJ...930...39V, 2018A&A...620A.128V}, and references therein). We adjust the compiled luminosities from previous works to the Gaia DR3 \citet{2021AJ....161..147B}  geometric distances. We then use these luminosities and effective temperatures to re-derive stellar masses homogeneously for the whole sample. For this, we use the PARSEC 1.2S pre-main sequence tracks (\citealp{2012MNRAS.427..127B}). A Hertzsprung-Russell (HR) diagram of the sample is shown in Fig.~\ref{Plot: The sample}. The result is 317 sources in the sample with stellar mass determinations. We did not derive stellar masses for the 20 remaining sources, whose effective temperatures and luminosities place them to the left of the theoretical pre-main sequence in the HR diagram (Fig.~\ref{Plot: The sample}, this is often caused by unresolved binarity). 




\subsection{Sample completeness and biases}

The intermediate to high-mass YSO regime comprises different historical types of sources. Between $1.5$\textless{}M\textless{}$8$ M$_{\odot}$, YSOs are categorized into the Herbig Ae/Be group (e.g. \citealp{2018A&A...620A.128V}), at the latest stages of pre-main sequence evolution, and their cooler predecessors the Intermediate Mass T Tauris (IMTTs, e.g. \citealp{2021A&A...652A.133V}). At M\textgreater{}$8$ to $10$ M$_{\odot}$, YSOs are generally referred to as Massive Young Stellar Objects (MYSOs, e.g. \citealp{2021A&A...654A.109K,2021A&A...652A..68M}). In this work, we use the term `intermediate and high-mass' when referring to any YSO with M\textgreater{}$1.5$ M$_{\odot}$ and optically bright enough to be detected by Gaia (as opposed to the T Tauri regime containing optically bright low-mass YSOs). If we refer to particular stellar mass ranges, we explicitly mention it in the text. We note that most of the sources in the sample fit within the classical Herbig Ae/Be regime, as it is the one more accessible with Gaia (Fig.~\ref{Plot: The sample}, c.f. \citealp{2023SSRv..219....7B}).

The sample of 337 stars considered in this work contains most of the known intermediate and high-mass YSOs which are well characterized. We note that there are thousands of other proposed intermediate and high-mass YSO candidates (e.g. \citealp{2008AJ....136.2413R, 2020A&A...638A..21V, 2023A&A...674A..26C}), but these lack spectroscopical confirmation of their nature and stellar parameters. We note that accurate stellar parameters and extinctions are necessary to derive stellar masses. For this reason, and the larger uncertainty on their true nature, we chose to leave these candidates out of this study. We thus expect the contamination of our sample to be very low. We note that the Gaia DR3 Apsis from DPAC was highly inefficient when identifying intermediate and high-mass YSOs (mainly because they are a very small fraction of the Gaia Universe DPAC addresses, see Ae/Be stars in Table 2 of \citealp{2023A&A...674A..28F}). Hence, no reliable sources can be included from Gaia DR3 Apsis' catalogues of young stars. There are a few other spectroscopically characterized intermediate-mass YSOs in the literature, which we have not considered to keep the sample as homogenous as possible (this includes some or all of the sources from \citealp{2021AJ....162..153N, 2021RAA....21..288S,2021A&A...652A.133V, 2023AJ....165....3K, 2022ApJS..259...38Z, 2023MNRAS.519.3958I}). We estimate that, within the distance range considered in the analysis and discussion of this work (350 to 4000 pc, Sect.~\ref{Sect. HDBSCAn results} and Fig.~\ref{Plot: The sample distance}), our sample contains at least 85\% of the known and spectroscopically characterized intermediate and high-mass YSOs.

We have not considered the YSOs which are very embedded in their parental material, and are not optically visible due to large extinctions (i.e., those who do not appear in Gaia, which has a faint limit of $G \lesssim 21$ mag). This implies we are mainly considering Class II and Class III YSOs in this study. Figure~\ref{Plot: The sample} shows an HR diagram and the stellar mass distribution of the sample. Looking at Fig.~\ref{Plot: The sample}, we see that our sample is covering well the range of 1.5 to 20 M$_{\odot}$. In principle, we would expect the above incompleteness to mainly affect the higher-mass regime, due to the faster evolution of more massive stars (e.g., Fig.~\ref{Plot: The sample} isochrones show that above $4$ M$_{\odot}$ all sources are younger than 1 Myr). However, fitting an Initial Mass Function (IMF, Fig.~\ref{Plot: The sample}) to the $2-3$ M$_{\odot}$ bin shows that the more massive population of YSOs is overrepresented in our sample (M\textgreater{}$8$ M$_{\odot}$). This is probably due to the fact that, although optically visible M\textgreater{}$8$ M$_{\odot}$ YSOs are rarer in their mass range, they can be observed at much larger distances. Therefore, our selected sample of sources can be considered as representative in mass of the evolved (Class II, Class III) Galactic population of intermediate and high-mass YSOs.

In the YSO massive regime ($M>8$ M$_{\odot}$), we have estimated that only $\sim20$\% of the total population appears in Gaia (comparing with the RMS survey, \citealp{2013ApJS..208...11L}). However, due to the fast evolution of these massive sources, it is uncertain whether Gaia is tracing the more evolved population or the population that shredded its envelope through some unknown dynamical process. In any case, the dynamical cluster dissipation timescales are much larger than the massive star formation timescales (\citealp{2023MNRAS.523.2083F,2023RAA....23g5023H}). Hence, this sample bias caused by the optical limitation of Gaia is likely to have a limited impact on our analysis of the clustering properties of intermediate and high-mass YSOs. A more detailed and technical analysis of the biases and completeness of the methodology is presented in Appendix~\ref{Appendix_B}.


In conclusion, the sample of 337 sources gathered in this work is not volume complete. However, it contains the majority of known and well-characterized intermediate and high-mass YSOs with accurate Gaia data, and it is representative in mass of the optically visible ($G \lesssim 21$ mag) YSO population ranging 1.5 to 20 M$_{\odot}$. More in depth information of the properties of this sample can be found in \citet{2015MNRAS.453..976F, 2018A&A...620A.128V, 2019AJ....157..159A, 2020MNRAS.493..234W, 2021A&A...650A.182G, 2022ApJ...926..229G}, and \citet{2022ApJ...930...39V}, among other works. A complete review of this population of objects, with references, is provided by \citet{2023SSRv..219....7B}.



\section{HDBSCAN clustering methodology}\label{S_HDBSCAN}

In this section, we use the unsupervised machine learning algorithm HDBSCAN (Hierarchical Density-Based Spatial Clustering of Applications with Noise, \citealp{10.1007/978-3-642-37456-2_14,mcinnes2017hdbscan}) to identify clusters and associations in the Gaia catalogue containing the intermediate and high-mass YSOs of the sample described in Sect.~\ref{S_sample}. \citet{2021A&A...646A.104H} concluded that HDBSCAN is the most sensitive and effective algorithm for recovering open clusters in Gaia data. \citet{2019A&A...626A..80C} got to the same conclusion identifying members of the young ($<5$ Myr) Rho Oph star forming region. We refer the reader to the aforementioned references for a detailed explanation on how HDBSCAN selects clusters. In the following subsections we describe our use of the Python implementation of HDBSCAN\footnote{\href{https://hdbscan.readthedocs.io/en/latest/how_hdbscan_works.html}{https://hdbscan.readthedocs.io/en/latest/how\_hdbscan\_works.html}}.


\subsection{HDBSCAN methodology}\label{S_s_HDBSCAN}

HDBSCAN has three hyperparameters of importance (i.e., algorithm related variables that need to be decided by the user). The hyperparameter `cluster\textunderscore selection\textunderscore method' defines how clusters are selected from the cluster tree. We use the `leaf' method in this work, as recommended by \citet{2021A&A...646A.104H} for identifying clusters in Gaia data. An additional advantage of `leaf' for this work is that it favours the selection of all the clusters present in a field, down to the smaller ones, without excluding the possibility of detecting large clusters. The two other hyperparameters to consider are `min\textunderscore cluster\textunderscore size' and `min\textunderscore samples'. The first one defines the smallest sample size we can consider a `cluster'. The second one can be understood as a quantifier of how conservative HDBSCAN is in selecting clusters. In this work, we set these two parameters to be equal unless stated otherwise, as it is customary in most HDBSCAN applications where there is no previous knowledge of the type and number of clusters present in the search field.

In order to select the field of stars to search for clusters, we queried Gaia DR3 for the HEALPix pixel level $5$ (Hierarchical Equal Area isoLatitude Pixelization, see \citealp{2005ApJ...622..759G}) which contains each of the massive YSOs, and the eight HEALPix pixels around it. Each searched area has hence $\sim31$ deg\textsuperscript{2} and side length approximately $5$ deg. According to \citet{2021A&A...646A.104H}, this angular size is sufficient for detecting clusters at the wide range of distances considered. The only quality requirement is that sources must have a 5-parameter Gaia DR3 astrometric solution (\textalpha{}, \textdelta{}, $\varpi$, \textmu{}\textsubscript{\textalpha{}}\textsubscript{*}, \textmu{}\textsubscript{\textdelta{}}, faint limit $G\sim21$ mag; \citealp{2021A&A...649A...2L}). No other quality constraint to the astrometry was applied. The effects of this, and the selection biases induced by demanding a Gaia DR3 5-parameter astrometric solution are proved to be very minor in Appendix~\ref{S_IR_images}. For 82 sources we used a smaller  HEALPix pixel as the fields were very crowded (level 6, $\sim8$ deg\textsuperscript{2} and side length of $2.5$ deg) and HDBSCAN was failing to identify similar scale-size associations. This smaller field size does not bias the cluster identification (see Appendix~\ref{Distance_effects} for a more detailed analysis). 



As a preprocessing step, we applied  the zero point bias correction to $\varpi$ as described in \citet{2021A&A...649A...4L}. In addition to this, the five dimensions of each field (\textalpha{}, \textdelta{}, $\varpi$, \textmu{}\textsubscript{\textalpha{}}\textsubscript{*}, \textmu{}\textsubscript{\textdelta{}}) were re-scaled to have a median of zero and a unit interquartile range. This re-scaling process ensures that each parameter has an equal weight for HDBSCAN (\citealp{2021A&A...646A.104H}).

We then applied the HDBSCAN algorithm to all Gaia DR3 stars in these fields using different combinations of parameters and hyperparameters. First, we looked for clusters in the 5 dimensional astrometric space (\textalpha{}, \textdelta{}, $\varpi$, \textmu{}\textsubscript{\textalpha{}}\textsubscript{*}, \textmu{}\textsubscript{\textdelta{}}). We did this twice, using `min\textunderscore cluster\textunderscore size' and `min\textunderscore samples' equaling $[10,10]$ and $[30,30]$. Then, we repeated the search for clusters in these fields in a 3 dimensional astrometric space (\textalpha{}, \textdelta{}, $\varpi$; `3d physical space'). In this latter case, we also ran the code twice, for `min\textunderscore cluster\textunderscore size' and `min\textunderscore samples' equaling $[10,10]$ and $[30,30]$. We thus ran HDBSCAN four times for each field containing each intermediate and high-mass YSO. This is done to cover the different combinations of possible cluster types (in physical space and proper motions or only in physical space) and minimum cluster sizes (10 and 30). An exploration of other HDBSCAN hyperparameters is presented in Appendix~\ref{Appendix_A}. This exploration shows that the hyperparameters chosen in this work are appropriate for identifying clusters and associations with HDBSCAN.

We note that circumstellar extinction is often significant in YSOs. This extinction varies from source to source and can only be characterized with spectroscopic data. This has prevented us from using the color-magnitude diagram as a tool to identify or evaluate clusters.

\begin{figure}[t!]
\includegraphics[width=\columnwidth]{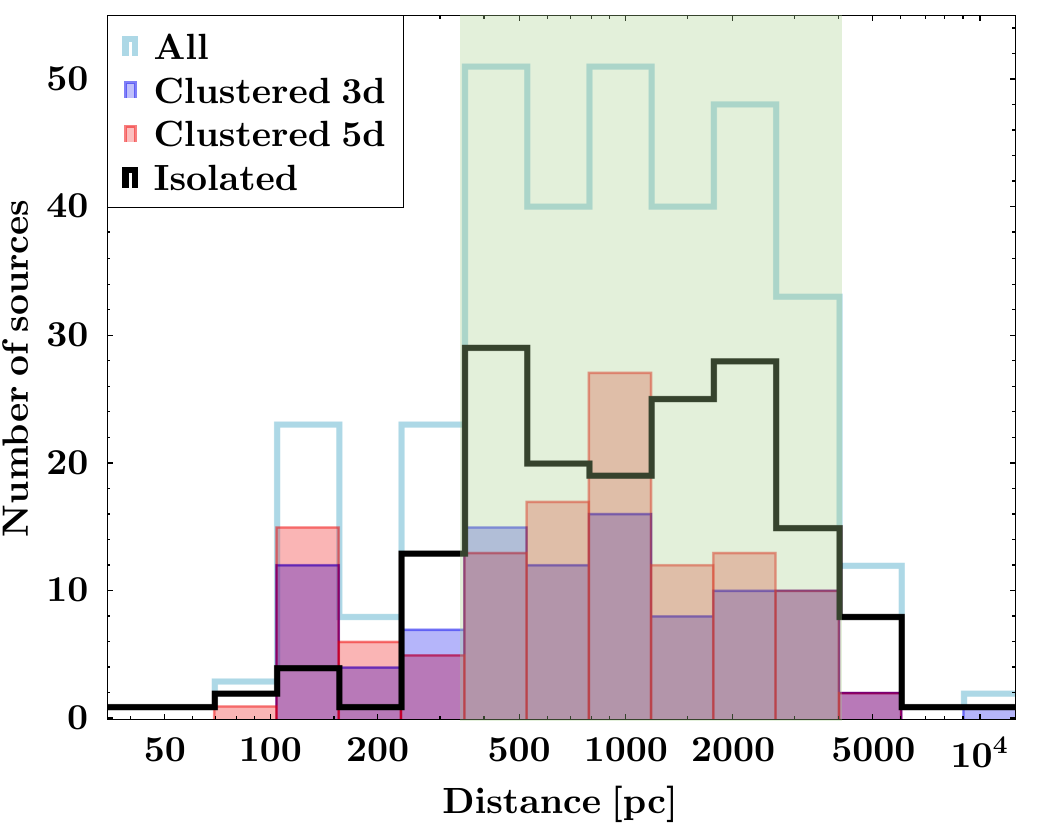}
\caption{Histogram of the 337 YSOs considered in this work as a function of distance (light blue contour). Black contours are isolated stars. Blue and red bars indicate stars clustered in 3d or 5d space, respectively (we note 49 sources appear clustered in both). The green area shows the region within which the clustering algorithm HDBSCAN is most sensitive, and where we limit our analysis. Distances are \citet{2021AJ....161..147B}  geometric distances.}\label{Plot: The sample distance}
\end{figure} 

\subsection{Results of HDBSCAN}\label{Sect. HDBSCAn results}


The methodology of the previous section was applied homogeneously to all 337 intermediate and high-mass YSOs in the sample. HDBSCAN assigns a normalized probability to every source in each field of belonging to a cluster or association. A probability of $0$ means the star is not in a cluster, while non-$0$ probabilities indicate different degrees of association to a cluster. To avoid biasing our results towards YSOs at cluster centres (which typically receive higher HDBSCAN probabilities), we considered as clustered the sources with a non-$0$ probability in either of the two configurations of `min\textunderscore cluster\textunderscore size' and `min\textunderscore samples' (higher probability threshold are considered in Sect.~\ref{S_Analysis} to support the analysis). As a result, we obtain 121/337 sources that appear as clustered in 5d (physical space and proper motions), and 97/337 stars that appear as clustered in 3d (physical space only). There is not a large overlap between both groups, with 49 sources appearing clustered in both parameter spaces. This is not surprising, all dimensions are treated equally by Sect.~\ref{S_s_HDBSCAN} methodology, and thus the proper motions make a significant difference in the clusters that can be identified or rejected (see Sect.~\ref{S_cluster_size} for a comparison between 3d and 5d clusters). All clusters have an HDBSCAN probability larger than 0.5, and 70\% of the detected clusters have an associated HDBSCAN probability larger than 0.95. There are 168 sources (50\% of the sample) that do not appear clustered in any HDBSCAN configuration, and hence they can be considered as isolated. 



These results of HDBSCAN are shown in Fig.~\ref{Plot: The sample distance} as a function of distance. From \citet{2021A&A...646A.104H}, we know the field angular size we are using is inefficient for detecting real associations below 350 pc. Fig.~\ref{Plot: The sample distance} confirms this. Likewise, it is evident from Fig.~\ref{Plot: The sample distance} that 4000 pc is likely a threshold for the HDBSCAN cluster identification sensitivity. Therefore, we define 350 to 4000 pc as the region where our HDBSCAN methodology is most sensitive, and unless otherwise specified we limit our results and analysis to it. There are both clustered and isolated stars throughout this distance range, containing 263 stars. 



Of the 263 stars, 92 are clustered in 5d (35\%), 71 are clustered in 3d (27\%), 36 are clustered in both 3d and 5d (14\%), and 127 are clustered in 3d or 5d (48\%). A total of 136 sources are isolated (52\%), not appearing clustered in any HDBSCAN configuration. The clustering nature, cluster properties, and HDBSCAN results for each intermediate and high-mass YSO considered in this work between 350 to 4000 pc are summarised in Table~\ref{Table}. 


\subsection{Cluster sizes and number of cluster members}\label{S_cluster_size}

\begin{figure*}[t!]
\includegraphics[width=\textwidth]{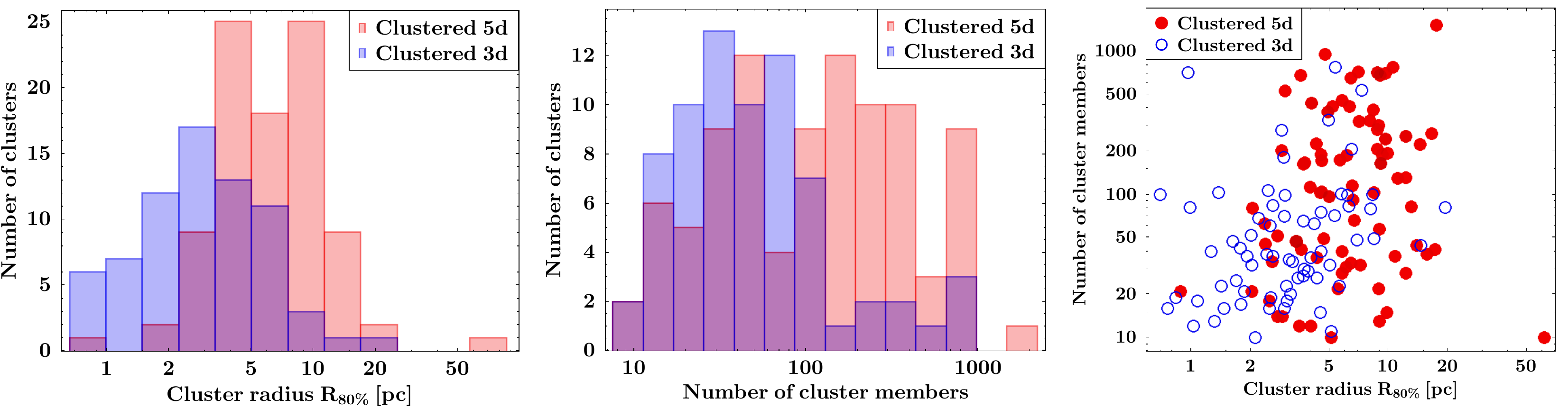}
\caption{\textit{Left:} Histogram of cluster radii identified in 3d and 5d spaces. Cluster radius is defined as that containing 80\% of the cluster's stars from the cluster geometrical centre. \textit{Center:} Histogram of the number of sources in each cluster. \textit{Right:} Cluster radii vs. number of sources per cluster identified in 3d (blue) and 5d (red) spaces. }\label{Plot: Cluster_size}
\end{figure*} 

In this section, we describe the properties of the identified clusters and associations. In Fig.~\ref{Plot: Cluster_size}, the sizes of all identified clusters and the number of sources they contain are shown. To characterize the cluster size, we used a radius defined as the angular distance to the geometrical cluster centre (in \textit{ra} \& \textit{dec}) that contains 80\% of the cluster's stars. The angular radius was converted to physical radius ($R_{80\%}$) using the distance to the massive YSO contained in each cluster. A radius containing 80\% of the clusters' stars was used instead of the more typical 50\% value (\citealp{2023A&A...673A.114H}) to better characterize the full extent of these clusters and associations (e.g. \citealp{2021A&A...645A..84M}). We note that some clusters have been identified using both a `min\textunderscore cluster\textunderscore size'  of 10 and 30 (Sect.~\ref{S_s_HDBSCAN}). In those cases, we always used as reference the more populated cluster identified in each case (as it has more statistical meaning). We also note that in some cases HDBSCAN identifies clusters with tidal tails, or that are heavily non-spherical, so the $R_{80\%}$ radii should always be interpreted as the size of the clusters in their more elongated direction.

Fig.~\ref{Plot: Cluster_size} shows that the clusters identified in 5d space are larger on average than those identified in 3d space, with $R_{80\%}$ mean and standard deviation of $7.6\pm6.8$ and $3.7\pm3.0$ pc, respectively. These differences in radial sizes are caused by HDBSCAN often being able to identify larger clusters when proper motion information is also provided. The large standard deviations show the large scatter in cluster sizes in both 3d and 5d. Overall, 90\% of the identified clusters are smaller than 10 pc, and all but one are smaller than 20 pc. Fig.~\ref{Plot: Cluster_size} also shows that the clusters identified in 5d space are typically more populated than those identified in 3d space, but again the spread is very large in either case. In 5d space, 45\% of the clusters have less than 100 members, but this percentage is of 80\% for the 3d case. Only 13\% and 6\% of clusters in 5d and 3d have more than 500 members, respectively. 

From Fig.~\ref{Plot: Cluster_size}, only two cases of cluster miss-identification can be suspected. The first one is the cluster with $R_{80\%}=62$ pc and 10 members. The second one is the cluster with $706$ members in less than a parsec. These two clusters might not be real associations, but we cannot rule out their existence. For consistency, we kept them in the analysis of Sect.~\ref{S_Analysis}, but in Table~\ref{Table} they are marked as possible contaminants.

There are 24 YSOs that share a cluster with other YSOs of our list. None of these sources are likely to be binaries considering their Gaia astrometry, and they span the whole extent of stellar masses considered (Fig.~\ref{Plot: The sample}). We find that these clusters with more than one intermediate and high-mass YSO are on the denser and more populated end of the cluster distribution (Fig.~\ref{Plot: Cluster_size}).

The population of known open clusters (see Fig. 12 of \citealp{2023A&A...673A.114H}) have $R_{50\%}$ typical radii of 1.5 to 5 pc, and up to 15 pc. These sizes are very similar to the ones we find in our population of clusters and associations. The number of cluster members in the population of known open clusters also match well the results we retrieve for our sample (\citealp{2023A&A...673A.114H}, see their Fig. 2), even for the more populated clusters we obtain. Hence, we conclude that the clusters and associations identified in this work are similar in size and in number of members to the population of known open clusters (see Sect.~\ref{S_lit_openc} for further comparison). 



\section{Analysis}\label{S_Analysis}

In this section, we analyse the clustered and isolated intermediate and high-mass YSOs identified in Sect.~\ref{S_HDBSCAN}, and their relation to measured cluster properties.


From here onwards, we define as `clustered' those YSOs appearing clustered in 3d or in 5d (Sect.~\ref{Sect. HDBSCAn results}), unless stated otherwise. We note that this definition considers clustered whatever star that is not clearly isolated in Gaia space, when the literature usually defines clustered sources in the opposite way (as stars clearly belonging to a group, see \citealp{2020A&A...640A...1C, 2023A&A...673A.114H}, and references therein). This approach was taken because Sects.~\ref{Sect. HDBSCAn results} and~\ref{S_cluster_size} have evidenced that, with our methodology, we are detecting different types of clusters in 3d and 5d. In Sect.~\ref{S_lit_openc} we compare the results of our approach with the population of known open clusters in the Galaxy.


\subsection{Clustered nature as a function of YSO mass}\label{S_Clus_vs_mass}

\begin{figure*}[t!]
\includegraphics[width=\textwidth]{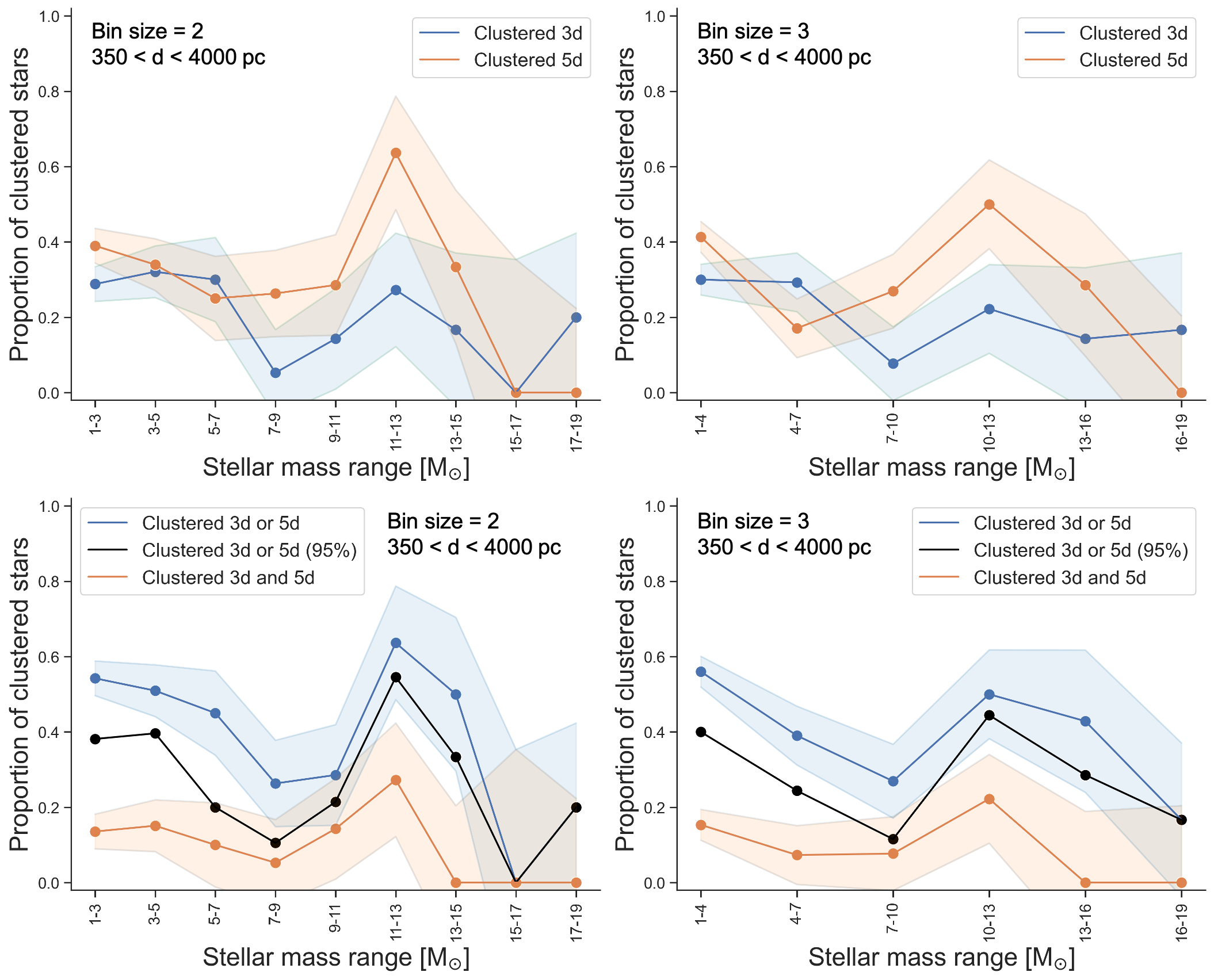}
\caption{Proportion of clustered intermediate and high-mass YSOs as a function of stellar mass. Top plots show the proportion of clustered stars in 3d and 5d HDBSCAN executions, left plot with 2 M$_{\odot}$ bins and right plot with 3 M$_{\odot}$ bins. Bottom plots show the proportion of clustered stars in either `3d or 5d' and both `3d and 5d'. In black, the `3d or 5d' results when only considering those sources for which HDBSCAN reports more than a 95\% clustering probability. Unless otherwise specified, in this work we define as `clustered' the `3d or 5d' combination (Sect.~\ref{S_Analysis}). Complete error is $1/\sqrt{n}$, with $n$ being the number of sources per bin.}\label{Plot: Mosaic 1}
\end{figure*} 

In Section~\ref{Sect. HDBSCAn results}, we concluded that 48\% of the known Galactic population of intermediate and high-mass YSOs appears clustered. In Fig.~\ref{Plot: Mosaic 1} we show the fraction of clustered intermediate and high-mass YSOs as a function of stellar mass. To minimize the impact of statistical flukes on the interpretation, different binning to the stellar mass were considered, as well as the four combinations of 3d, 5d, `3d and 5d', and `3d or 5d' HDBSCAN results (Sect.~\ref{Sect. HDBSCAn results}).



Regarding the `3d or 5d' case, Fig.~\ref{Plot: Mosaic 1} shows that at the lower-mass end of the sample ($1.5-4$ M$_{\odot}$, mainly the Herbig Ae and IMTT regimes) $\sim55\%$ of the sources are clustered. This number nicely matches the proportion of clustered stars found in the T Tauri regime using similar methodologies (e.g. \citealp{2020AJ....160...68W,2021ApJS..254...33K}). However, as we go up in stellar mass ($4$ to $10$ M$_{\odot}$), the proportion of clustered stars diminishes, down to $\sim27\%$ at the $7-10$ M$_{\odot}$ range. The proportion of clustered stars goes up again from $10$ M$_{\odot}$ to a $\sim65\%$ clustering rate at $11-13$ M$_{\odot}$, but it seems to decay abruptly shortly after that, and most stars above $15$ M$_{\odot}$ appear isolated (albeit see below and Sect.~\ref{S_dis_more_massive}). This trend of intermediate-mass YSOs in the range $4$ to 10 M$_{\odot}$ being less clustered than lower-mass YSOs ($<4$ M$_{\odot}$) and the massive YSOs ($10-13$ M$_{\odot}$) is noticeable in all other HDBSCAN results (3d, 5d, 3d and 5d, Fig.~\ref{Plot: Mosaic 1}) and mass binning. Furthermore, this trend is also apparent when only considering as clustered those sources for which HDBSCAN reports more than a 95\% clustering probability (Sect.~\ref{Sect. HDBSCAn results}, black lines of Fig.~\ref{Plot: Mosaic 1}).

We perform a two-sample Kolmogorov–Smirnov (KS) test to probe the significance of these variations in clustering rate as a function of stellar mass. The null hypothesis is that, in stellar mass, the groups of clustered and isolated YSOs are drawn from the same unknown probability distribution. When the KS test is performed using the whole sample, we obtain a p-value of 0.025.  Because above 10 M$_{\odot}$ we are affected by low number statistics and some biases (see below and Sect.~\ref{S_dis_more_massive}), we also report a p-value of 0.041 when we limit the analysis to 10 M$_{\odot}$. We repeat this KS analysis only considering as clustered those sources for which HDBSCAN reports more than a 95\% clustering probability (Sect.~\ref{Sect. HDBSCAn results}). The p-values obtained in this latter case are 0.083 (whole sample) and 0.039 (limited to 10 M$_{\odot}$). We conclude that these KS tests show that the clustering rate varies with stellar mass to within $\sim95$\% significance. 



\begin{figure}[t!]
\includegraphics[width=\columnwidth]{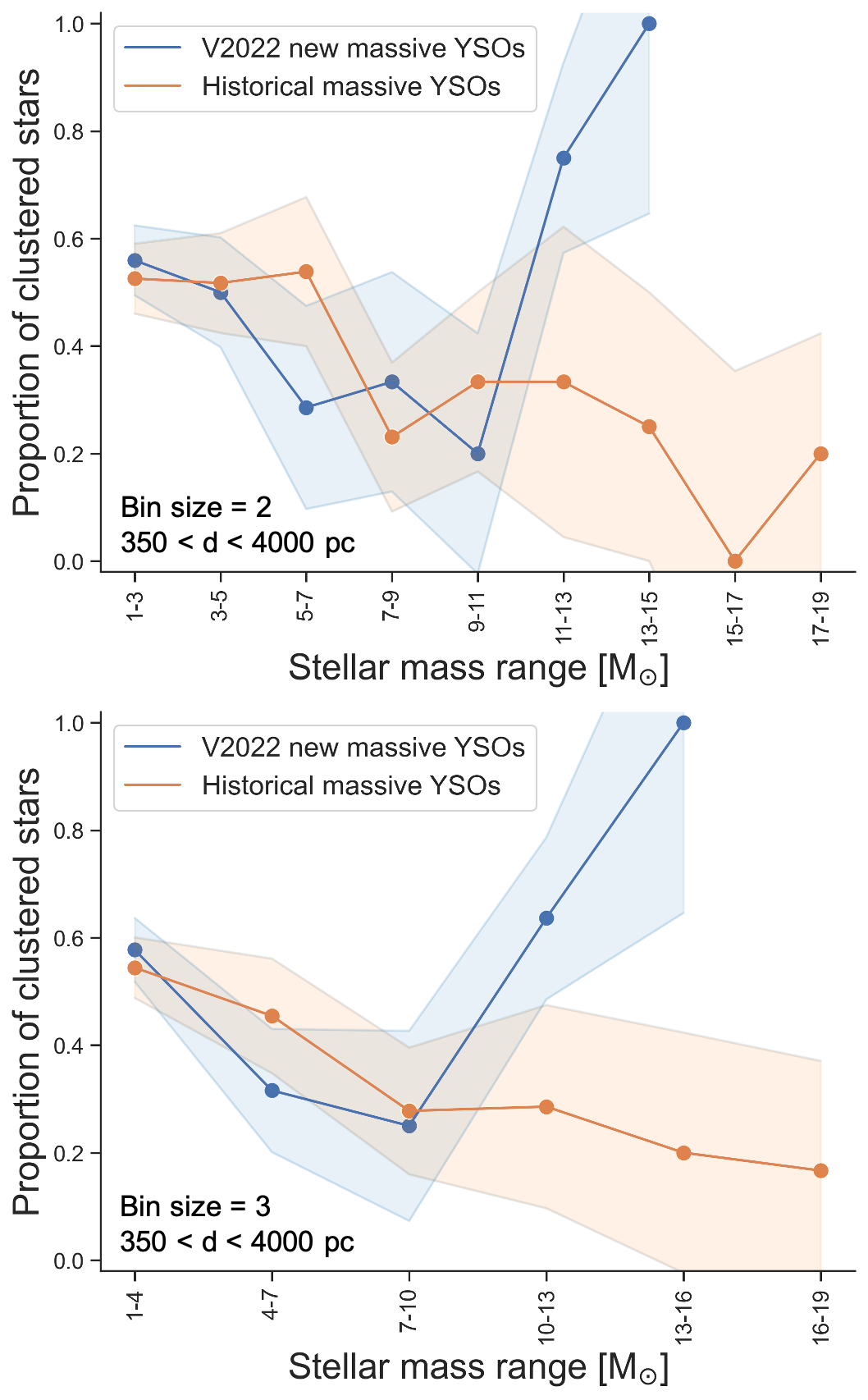}
\caption{Proportion of clustered (`3d or 5d') intermediate and high-mass YSOs as a function of stellar mass. Historically known YSOs have been separated from those identified homogeneously and in a position-unbiased way by \citet{2022ApJ...930...39V}. Top and bottom plot differ in the stellar mass bin used. Complete error is $1/\sqrt{n}$, with n being the number of sources per bin.}\label{Plot: Mosaic 2}
\end{figure}

As described in Sect.~\ref{S_sample}, the considered sample of YSOs has two main components, a heterogeneous collection of historically known and studied intermediate and high-mass YSOs (compiled from \citealp{2018A&A...620A.128V} and \citealp{2021A&A...650A.182G}), and a set of intermediate and high-mass YSOs homogeneously identified with all-sky surveys (\citealp{2020A&A...638A..21V,2022ApJ...930...39V}). In Fig.~\ref{Plot: Mosaic 2}, we show the clustering ratio as a function of stellar mass of these two groups. It is noteworthy that, although the trend of clustering ratio going down with stellar mass until $10$ M$_{\odot}$ holds for both samples, the two sets show opposed behaviours above $10$ M$_{\odot}$. For masses $M>10$ M$_{\odot}$, the newly identified sources show a sharp increase of the clustering rate up to a $60-100$\% clustering rate at $15$ M$_{\odot}$, whereas the historically known sources show a smooth decline of their clustering rate with mass, with only $0-40$\% of them above 10 M$_{\odot}$ appearing clustered. This trend is also apparent when only considering as clustered those sources for which HDBSCAN reports more than a 95\% clustering probability (Sect.~\ref{Sect. HDBSCAn results}), or when considering only HDBSCAN results in 3d, 5d, and `3d and 5d'. We note that the sample from \citet{2022ApJ...930...39V} is limited to $15$ M$_{\odot}$, whereas the historical sample extends to $20$ M$_{\odot}$. We also note that the low number of sources at these stellar mass ranges do not allow for robust statistics, as it is indicated by the large uncertainties of Fig.~\ref{Plot: Mosaic 2} (see Sect.~\ref{S_dis_more_massive} for further discussion).

\subsection{Cluster properties as a function of YSO mass}\label{Cluster properties as a function of YSO mass}

\begin{figure*}[t!]
\includegraphics[width=\textwidth]{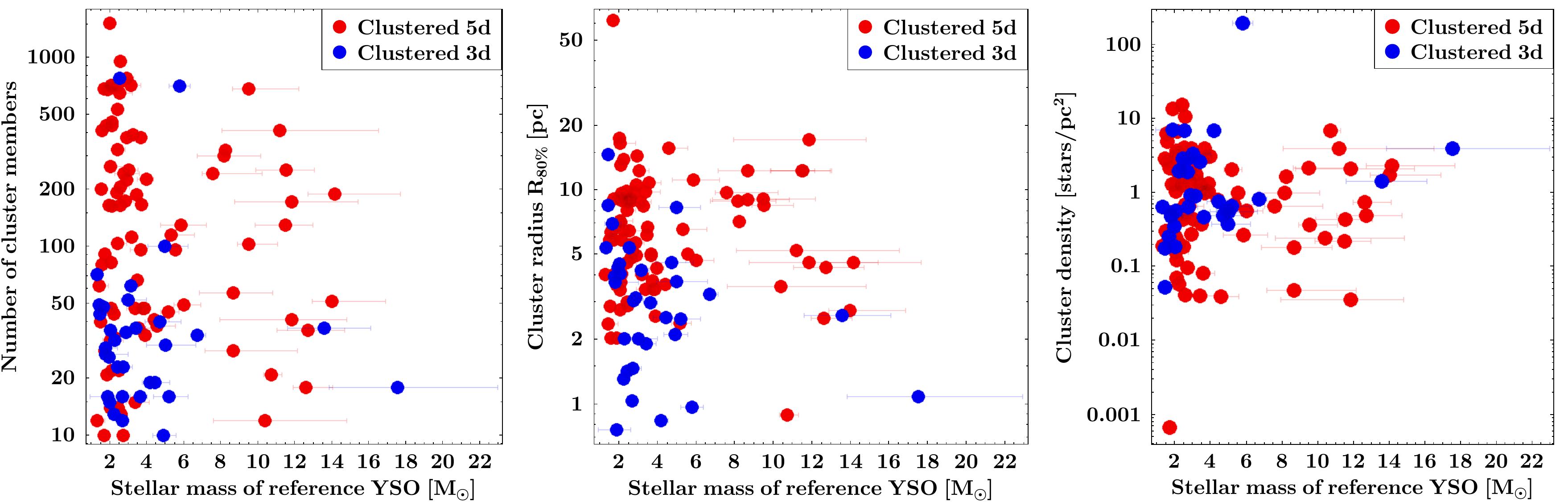}
\caption{Cluster properties as a function of the stellar mass of the intermediate or high-mass YSO contained in each cluster. \textit{Left:} Number of sources in each cluster ($N_{c}$). \textit{Center:} Cluster size ($R_{80\%}$). \textit{Right:} Cluster density ($0.8N_{c}/\pi R_{80\%}^2$).}\label{Plot: Cluster_analysis}
\end{figure*}

In this section, we study the properties of the identified clusters as a function of the properties of the intermediate and high-mass YSO they contain. Figure~\ref{Plot: Cluster_analysis} shows the YSO stellar masses as a function of the number of stars per cluster ($N_{c}$), the cluster radii ($R_{80\%}$), and the cluster densities. The cluster density is defined as $0.8N_{c}/\pi R_{80\%}^2$, and thus can be understood as the average number of cluster stars per square parsec, assuming the clusters have a circular distribution in the sky (which is not always a good assumption, Sect.~\ref{S_cluster_size}). In the case of the clusters that were identified both in 3d and in 5d, in Figure~\ref{Plot: Cluster_analysis} we report the properties of the 5d clusters, which typically have more members and statistical significance (see Fig.~\ref{Plot: Cluster_size}). Most clusters have 0.1 to 10 stars per squared parsec, with an average mean of $3.2$ stars/pc\textsuperscript{2}. As illustrated in Fig.~\ref{Plot: Cluster_analysis}, we find no dependence of cluster properties with the stellar mass of the considered intermediate or high-mass YSO.

\begin{figure}[t!]
\includegraphics[width=\columnwidth]{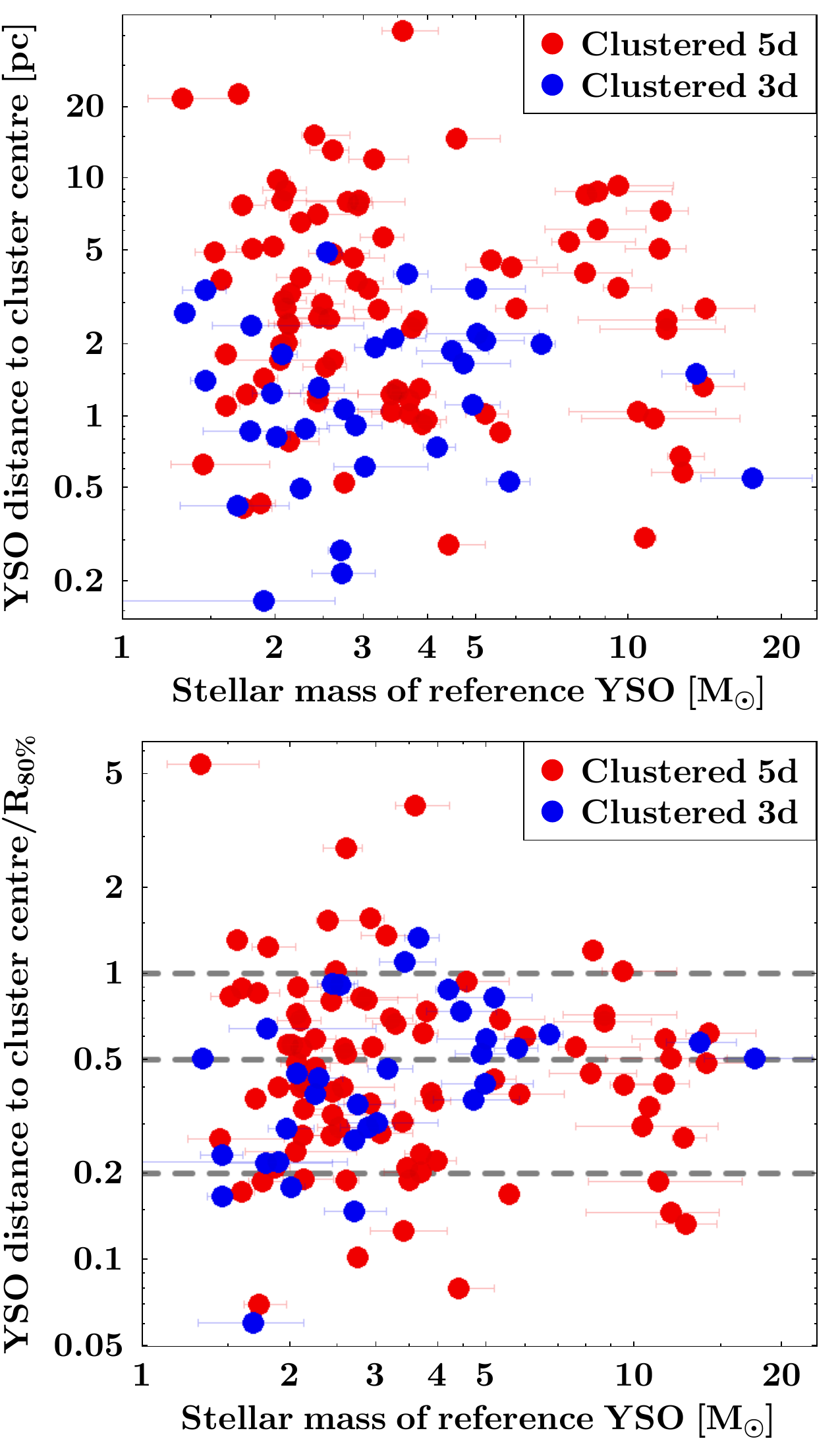}
\caption{\textit{Top:} Distances of the intermediate and high-mass YSOs to the geometrical centre of their clusters as a function of their stellar mass. \textit{Bottom:} Distances of the intermediate and high-mass YSOs to the geometrical centre of their clusters divided by the cluster sizes ($R_{80\%}$) vs. the stellar mass of the intermediate and high-mass YSOs.}\label{Plot: Cluster_analysis_2}
\end{figure}

We also study the location of the intermediate and high-mass YSOs within the clusters, to trace possible mass segregation. In Fig.~\ref{Plot: Cluster_analysis_2}, we show the angular distances from the intermediate and high-mass YSOs to the geometric centre of the clusters (in \textit{ra} \& \textit{dec}), converted to physical distances by using the distances to the YSOs. Again, we find no correlation between the stellar masses of the intermediate and high-mass YSOs and their distances to the cluster centres. The only appreciable fact is that above 6 M$_{\odot}$ no YSO is beyond 10 pc of its cluster centre, but there is no statistical significance in this given that most clusters are smaller than 10 pc (Sect.~\ref{S_cluster_size}, Fig.~\ref{Plot: Cluster_size}). In 11\% of the cases, the intermediate or high-mass YSO is located beyond the $R_{80\%}$ radius. This percentage goes up to 46\% and 86\% for $R_{80\%}/2$ and $R_{80\%}/5$ radii, respectively (gray lines of Fig.~\ref{Plot: Cluster_analysis_2}). Therefore, we estimate that only $\sim15\%$ of the clustered intermediate and high-mass YSOs are in the inner 20\% region of the clusters, and that this percentage does not depend much on the stellar mass of the YSO (Fig.~\ref{Plot: Cluster_analysis_2}). Again, only a weak trend can be reported, which is that the intermediate and high-mass YSOs under 4 M$_{\odot}$ mostly appear as the only ones outside the $R_{80\%}$ radius.

\subsection{Comparison with literature open clusters and associations}\label{S_lit_openc}

In this section, we compare the clusters obtained with HDBSCAN in Section~\ref{S_HDBSCAN} with different compilations of open clusters and associations of the literature, most of them also based in Gaia DR3. We note that this comparison is limited to the 263 sources within 350 to 4000 pc. Unless otherwise specified, cross-matches with other catalogues were done with a 5 arcsecond aperture.

\begin{itemize}
    \item \citet{2023A&A...673A.114H}. We have 78 stars in common with this catalogue of cluster members. Of these 78, we found 70 clustered and 8 isolated.     
    \item \citet{2023arXiv230608150K}. There are 41 stars in our sample which the SPYGLASS Catalog of Young Associations identified as clustered (P\textsubscript{mem}$>0.95$). Of these 41, we identified 28 as clustered.
    \item \citet{2022A&A...661A.118C}. We have only one star in common with this catalogue of cluster members, and we also identify it in a cluster.
    \item \citet{2022A&A...664A.175P}. This is an unbiased survey of YSO associations in the Milky Way. Only two stars of our YSO sample match their list of members. Of these two, one we detect as clustered and one as isolated.
    \item \citet{2021MNRAS.504..356D}. Our sample of YSO sources has six stars matching this catalogue of YSO clusters within 2 arcminutes. We also identify these six sources in a cluster.
    \item \citet{2020A&A...640A...1C}. We have 37 stars in common with this catalogue of cluster members. Of these 37, we have detected 30 as clustered.
    \item \citet{2020A&A...633A..99C}. We have 46 stars in common with this catalogue of cluster members. Of these 46, we have detected 35 as clustered.
    \item \citet{2019AJ....158..122K}. We have 50 stars in common with this catalogue of cluster members. Of these 50, we have detected 36 as clustered.
\end{itemize}

Regarding infrared based catalogues of clusters and associations:

\begin{itemize}
    \item SPICY catalogue of YSOs (\citealp{2021ApJS..254...33K}). There are only 24 YSOs of our list in SPICY. Of these, SPICY reports 15 as clustered, of which we find 11. In contrast, we find 14 clustered stars and 10 isolated, and 11 and 4 of them appear clustered in SPICY, respectively. We note that SPICY only considered clusters of suspected YSOs in its HDBSCAN implementation, and used different HDBSCAN hyperparameters to those of this work (see Sect.~\ref{S_s_HDBSCAN}). 
    \item \citet{2019AJ....157...12B}. We have 12 stars in common with this catalogue of cluster and association members. Of these 12, we have detected 7 as clustered.    
    \item \citet{2013A&A...558A..53K}. Our sample of YSO sources has 18 stars matching this catalogue of YSO clusters within 2 arcminutes. We identify 13 of these sources in a cluster.
    \item \citet{2009ApJS..184...18G}.  We have 9 stars in common with this catalogue of cluster members. Of these 9, we have detected 8 as clustered.
\end{itemize}

Hence, the general agreement with previous surveys of clusters and associations is of $70-90$\%. This agreement holds for IR-based cluster surveys (e.g. 73\% for SPICY and 89\% for \citealp{2009ApJS..184...18G}). If we only consider the most recent and complete survey looking for clusters (\citealp{2023A&A...673A.114H}), the agreement is of 90\%. When analysing the YSOs that we report as isolated and other authors reported as clustered, we do not see any trend favouring more massive objects. Hence, these comparisons show no bias in our methodology towards detecting less clusters at higher stellar masses.


The works of \citet{2023A&A...673A.114H}, \citet{2022A&A...661A.118C}, and \citet{2020A&A...640A...1C} claim to be almost complete within 2 kpc, where $\sim76\%$ of our YSOs are (Fig.~\ref{Plot: The sample distance}). It is hence noteworthy that 55 clustered stars in our analysis are not listed in these works. For example, 36 of the 94 clustered stars we have identified within 2000 pc do not appear clustered in \citet{2023A&A...673A.114H}. We believe this is due to the looser definition of `cluster' used in this work (Sect.~\ref{S_s_HDBSCAN}, Sect.~\ref{S_Analysis}). We retrieve most of the clusters reported in the literature, but also other less bound associations and stellar overdensities that were discarded by previous authors.

The main differences between our work and most other Gaia-based cluster works is that we use both a 3d and a 5d space to identify clusters, we do not attempt to fit the clusters photometrically in a color-magnitude diagram (because of circumstellar extinction and because e.g., sequential star formation might have occurred, \citealp{2005AJ....129..776N}), and we do not perform a post-HDBSCAN case-by-case vetting (in order to keep the results homogeneous, see Sect.~\ref{S_Analysis}). Indeed, an in-depth look shows that the clusters we report which \citet{2023A&A...673A.114H} did not identify are typically the less dense ones, with larger radii, and fewer members. In particular, many of the clusters reported here and unidentified by \citet{2023A&A...673A.114H} have less than 50 members. We report these 55 new clusters that we have found in Gaia space for future reference and analysis (Table~\ref{Table}).

We emphasise that our approach was designed to trace exhaustively and homogeneously the population of clustered intermediate and high-mass YSOs. The strength of our analysis is that our methodology has been applied homogeneously to all sources, without favouring any interpretation of the definition of `cluster', and that it includes both the 3d and 5d Gaia astrometric spaces for cluster selection. It is beyond the scope of this work to describe individual clusters, but we caution the reader that some of the clusters reported here can be loose or sparse associations, not fitting the canonical definition of open cluster. Indeed, many known YSO associations do not fit the definition of open cluster (e.g. the TW Hydrae sparse association, \citealp{2023AJ....165..269L}). We refer the reader to the works of  \citet{2023A&A...673A.114H}, \citet{2022A&A...661A.118C}, and \citet{2020A&A...640A...1C} for a description, analysis, and discussion of open cluster identification in the Gaia era.


\subsection{Comparison with Iglesias et al. 2022}\label{S_Vioque_Daniela}

\citet{2023MNRAS.519.3958I} selected 220 intermediate-mass YSOs ($1-7$ M$_{\odot}$) within 400 pc as traced by their IR excess at 22 \textmu{}m. It is interesting to compare \citet{2023MNRAS.519.3958I} sample with the one in this work because there are significant differences between them. Although both samples were selected by demanding a certain IR-excess level, our selection often resulted in the presence of emission lines (mostly tracing active accretion, e.g. \citealp{2022ApJ...930...39V,2023SSRv..219....7B}) whereas \citet{2023MNRAS.519.3958I} selection did not (only 6 of their 220 stars have detected emission lines). This indicates that the sample considered in this work, even though being representative in mass distribution of the IMF (Sect.~\ref{S_sample}), might be a subsample of the total Galactic population of intermediate and high-mass YSOs; i.e., the subsample of sources with larger accretion rates. The results of \citet{2023arXiv230811430G} seem to point in this direction. It is unclear whether \citet{2023MNRAS.519.3958I} sample is more evolved than the one compiled for this work (Sect.~\ref{S_sample}), or if we have been biased in the past towards mainly identifying YSOs undergoing accretion bursts (see Sect.~\ref{S_Discussion} for further discussion).

\begin{figure}[t!]
\includegraphics[width=\columnwidth]{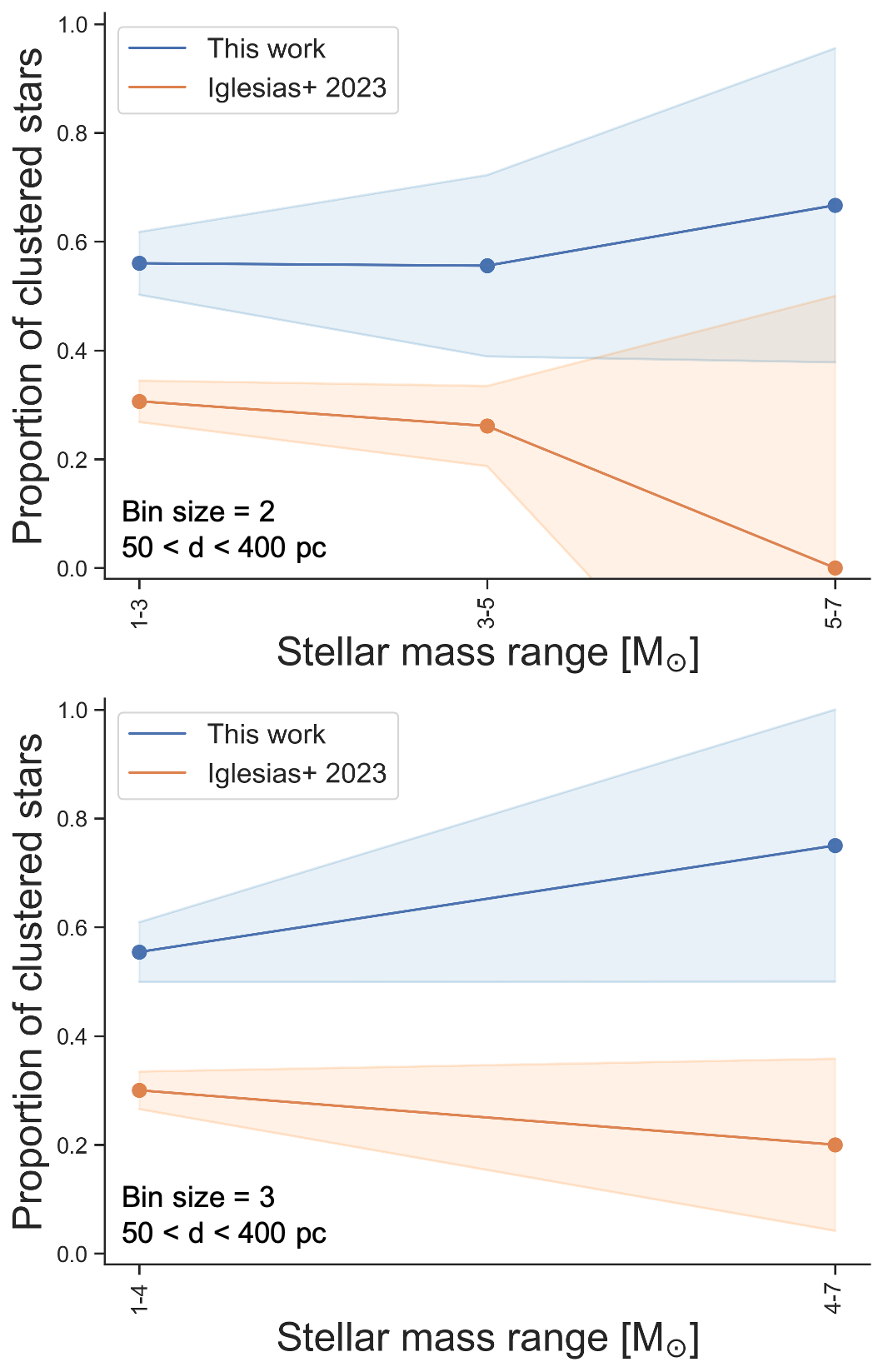}
\caption{Comparison of the clustering rate between the YSO sample of this work and the YSO sample of \citet{2023MNRAS.519.3958I}. Complete error is $1/\sqrt{n}$, with $n$ being the number of sources per bin. For visualization purposes, the plots show a smaller range of stellar masses with respect to Fig.~\ref{Plot: Mosaic 1}, as \citet{2023MNRAS.519.3958I} sample is limited to 7 M$_{\odot}$.}\label{Plot: Iglesias}
\end{figure}

In this section, we apply the exact same HDBSCAN methodology of Sect.~\ref{S_HDBSCAN} to the sample of intermediate-mass YSOs of \citet{2023MNRAS.519.3958I}. \citet{2023MNRAS.519.3958I} sample is contained within 400 pc. As described in Sect.~\ref{Sect. HDBSCAn results}, this short distance is suboptimal for the HDBSCAN methodology of this work. Hence, to avoid any possible bias in the comparison with our sample, we also limited our sample to 400 pc in this section.  The clustering rate as a function of stellar mass for both samples is shown in Fig.~\ref{Plot: Iglesias}. For every mass bin, the clustering rates of \citet{2023MNRAS.519.3958I} sample are lower. We report a $\sim60\%$ clustering rate for our sources, and a $\sim25\%$  clustering rate for \citet{2023MNRAS.519.3958I} sources. Both samples show roughly constant clustering rates as a function of stellar mass up to 5 M$_{\odot}$ (within error bars). Beyond 5 M$_{\odot}$ low number statistics impede us to reach meaningful conclusions.

In Appendix~\ref{Distance_effects}, we describe the detected biases of applying our clustering methodology (Sect.~\ref{S_HDBSCAN}) to sources at short distances. We find a bias against identifying sources as isolated below 200 pc (Fig.~\ref{Plot: Distance_effects_1}). We note that 57\% of \citet{2023MNRAS.519.3958I} sources are within 200 pc. In contrast, 38\% of the sources considered in this section from our sample are within 200 pc. Therefore, if correcting from this bias, we would expect to find an even larger difference in clustering rate between both samples.


\subsection{Biases and incompleteness}\label{S_bias_Summary}

In Appendix~\ref{Appendix_B} we study in depth the caveats, biases, and sources of incompleteness of the methodology and analyses described in Sects.~\ref{S_HDBSCAN} and~\ref{S_Analysis}. This section provides an informative summary of Appendix~\ref{Appendix_B} conclusions.

We have applied the same methodology homogeneously to most known and spectroscopically characterized intermediate and high-mass YSOs. According to the IMF, the considered sample of YSOs is representative in stellar mass of the population of intermediate and high-mass YSOs in the Galaxy ($1.5$-$20$ M$_{\odot}$, Fig.~\ref{Plot: The sample}). This is further reinforced by the fact that the sample is well-distributed in stellar mass across the Galaxy. Nevertheless, this sample has some selection biases. Mainly, we are limited to objects bright enough in the optical to be detected by Gaia (faint limit $G\sim21$ mag). Hence, we expect to be mostly tracing Class II and III objects, as the younger Class I sources are often too extincted for Gaia. This limitation also applies to the sources used to look for clusters with HDBSCAN, this resulting in incomplete Gaia 5-parameter (5d) fields.

We have analysed the Gaia 5d completeness by comparing with other optical and IR surveys and YSO catalogs. We conclude that some more stars per mass bin would have appeared clustered if there were Gaia data for all embedded sources around each YSO. Hence, the results presented in Figs.~\ref{Plot: Mosaic 1},~\ref{Plot: Mosaic 2}, and~\ref{Plot: Iglesias} are likely lower limits. However, we do not see any dependence with stellar mass in this limitation caused by Gaia 5d incompleteness. Hence, the evolution of clustering rate as a function of stellar mass is robust (i.e. the shape of Figs.~\ref{Plot: Mosaic 1},~\ref{Plot: Mosaic 2}, and~\ref{Plot: Iglesias}). We report no dependence of our results with the astrometric quality of the sources used for HDBSCAN.

We have also analysed whether there are distance or sky location biases affecting our analyses and conclusions. We find that none of our results depend on biases of that sort. In particular, we observe the trend of clustering rate going down in the mass regime $4-10$ M$_{\odot}$ at all distance ranges (Fig.~\ref{Plot: Distance_effects_2}). The reported cluster properties are also independent of the distance.

There is some ambiguity in the literature with respect to the definition of a `cluster'. In this work, we claim that a YSO is in a `cluster' if HDBSCAN detects it as in a stellar overdensity in the Gaia 3d or 5d astrometric spaces, with respect to the surrounding field. In Sect.~\ref{S_cluster_size}, we evidence that the clusters identified in this work have similar sizes and number of members to the population of open clusters identified with Gaia (c.f. \citealp{2023A&A...673A.114H}). In addition, we have identified most of the clusters reported in the literature containing the YSOs we have considered (Sect.~\ref{S_lit_openc}). This evidences that our clustering detection methodology is robust and consistent with previous efforts looking for clusters and associations. In addition, we have identified 55 previously unreported clusters. These new clusters are in general of a lower stellar density, larger, and with fewer members than most of the known open clusters. We thus believe that our methodology is more loose in the definition of cluster, and that these new clusters we have identified were likely discarded by previous authors' methodologies who aimed to achieve a high-purity in the canonical interpretation of open cluster (e.g. \citealp{2020A&A...640A...1C,2022A&A...661A.118C,2023A&A...673A.114H}). We note that there are several known sparse associations of YSOs that also do not fit the standard definition of open cluster, which are also typically larger and with fewer members (e.g. \citealp{2022AJ....164..151L,2023AJ....165..269L,2023AJ....165...37L}). We thus provide a more exhaustive identification of the population of clustered or associated intermediate and high-mass YSOs.




\section{Discussion}\label{S_Discussion}


In this work, we have used Gaia to study the clustering properties of the known and spectroscopically characterized population of intermediate and high-mass YSOs ($1.5$-$20$ M$_{\odot}$, Sect.~\ref{S_sample}). We have limited our analysis to the 350 to 4000 pc range. In this section, we discuss our results and the implications for star formation mechanisms at different stellar mass ranges.

\subsection{Clustering rate as a function of stellar mass}\label{Key_section}

Of the 263 stars considered in the analysis, 127 are clustered in 3d or 5d (48\%) and 136 sources are isolated (52\%), not appearing clustered in any HDBSCAN configuration. This fraction varies with stellar mass (Fig.~\ref{Plot: Mosaic 1}), and in Sect.~\ref{S_Clus_vs_mass} we prove this variation to be statistically significant. In particular, we find that the lower mass YSOs ($1.5$-$4$ M$_{\odot}$) have clustering rates of $\sim55\%$, a percentage similar to that found in the T Tauri regime (when a similar methodology and clustering tools are used). However, this clustering rate goes down with stellar mass in the $4$-$10$ M$_{\odot}$ regime, down to $7$-$10$ M$_{\odot}$ YSOs showing a $\sim27\%$ clustering rate. For sources in the $10-15$ M$_{\odot}$ regime, the clustering rate goes up again to a value similar or slightly higher than the one found in the lower-mass regime (see the discussion in next subsection). Although these percentages are likely lower limits due to optical extinction (Sect.~\ref{S_bias_Summary}), we note that this overall shape of the clustering rate as a function of stellar mass is independent of distance or Gaia completeness (Appendix~\ref{Appendix_B}), and hence we suspect it is an intrinsic property of star formation. In addition, this result is independent of how we choose to select our clusters (i.e., 3d, 5d, 3d and 5d, 3d or 5d, Sect.~\ref{S_HDBSCAN} and Fig.~\ref{Plot: Mosaic 1}). We hence conclude that intermediate-mass $4$-$10$ M$_{\odot}$ YSOs are typically less clustered than low $M<4$ M$_{\odot}$ and high $M>10-15$ M$_{\odot}$ mass YSOs  (see Sect.~\ref{S_dis_more_massive} for some caveats on the high-mass regime).

It is not surprising that at the low-mass end ($<4$ M$_{\odot}$) we see only $50-60\%$ of clustered stars. \citet{2023ASPC..534....1C} show that molecular clouds are very hierarchical and fractal systems, with short dynamical timescales. In addition, regions of filamentary collapse to hubs exist, but most stars do not form in hubs (only ~10\% do, \citealp{2023ASPC..534....1C}). Hence, many low-mass stars might have formed relatively isolated. In contrast, \citet{2014prpl.conf..291L} and \citet{2023ASPC..534...43Z} show a high-degree of clustered low-mass YSOs, suggesting that those low-mass YSOs appearing isolated might be the result of fast cluster dispersion (e.g. \citealp{2003ARA&A..41...57L, 2023RAA....23g5023H}) and ejection. In any case, many of the considered low-mass YSOs had time to be dispersed in the field.

However, it is noteworthy that for the intermediate-mass YSOs in the range $4-10$ M$_{\odot}$, with lifetimes one to two orders of magnitude shorter, the clustering rate goes down with stellar mass. One explanation could be that more massive YSOs form in more massive clusters, and that these tend to disperse faster (e.g. \citealp{2007MNRAS.376.1879W,2016A&A...590A.107O,2023MNRAS.523.2083F,2023RAA....23g5023H}). However, this would not explain why these intermediate-mass YSOs are often less clustered than the more massive YSOs above $10$ M$_{\odot}$. We suggest that intermediate-mass YSOs in the range $4-10$ M$_{\odot}$ have typically a more isolated formation than lower-mass YSOs, and that this is more the case for the more massive stars within this mass regime. To explain this, we put forward the idea that the most massive YSOs ($>10$ M$_{\odot}$) often form via a mechanism that demands that many stars are formed around them (like the competitive accretion mechanism, \citealp{2001MNRAS.323..785B, 2004MNRAS.349..735B}), producing their clustered nature. In contrast, intermediate-mass and low-mass YSOs typically form in an individual manner (e.g. via the monolithic collapse of an individual core, \citealp{2003ApJ...585..850M, 2004ApJ...603..383T}). However, intermediate-mass YSOs in the $4-10$ M$_{\odot}$ range do not appear often around very massive YSOs. This could be due to the sharp decrease of the Initial Mass Function, or due to some other effect. This would explain why low and intermediate-mass YSOs have different clustering rates, and why the clustering rate decreases with stellar mass in the intermediate-mass regime. 




\subsection{Clustering properties of the more massive YSOs ($M>10$ M$_{\odot}$) and evolutionary trends}\label{S_dis_more_massive}

Fig~\ref{Plot: Mosaic 1} shows that the clustering rate goes up from $10$ M$_{\odot}$ to $13$ or even $15$ M$_{\odot}$, but it is not trivial to describe the clustering rate of the more massive YSOs ($M>10$ M$_{\odot}$). Only 31 stars of our sample within 350 to 4000 pc have masses above $10$ M$_{\odot}$, and only 13 of these have masses above $13$ M$_{\odot}$. Hence, we warn the reader that the clustering rates reported for the more massive star formation regime ($>13$ M$_{\odot}$) might suffer from low-number statistics and unknown population biases. We note that this does not affect the results of Sect.~\ref{Key_section}, as it is the decrease in clustering rate as a function of stellar mass in the intermediate-mass regime ($4$-$10$ M$_{\odot}$) what drives the conclusions of that section.


In Sect.~\ref{S_Clus_vs_mass}, we noted that there is a difference in clustering properties when we divide the sample in those intermediate and high-mass YSOs identified homogeneously in \citet{2022ApJ...930...39V} and those YSOs historically known and compiled from the literature. The difference only appears from 10 M$_{\odot}$ (Fig.~\ref{Plot: Mosaic 2}). For those sources from \citet{2022ApJ...930...39V} there is a sharp increase of the clustering rate (up to 100\% within error bars at 15 M$_{\odot}$, where the sample ends), whereas for the historically known sources there is a monotonic decrease. We report no difference in the clusters identified for both sets, and there is no correlation between cluster properties and stellar mass in either sample (see Sect.~\ref{Cluster properties as a function of YSO mass}). In addition to the way they were identified, the only difference between both sets of sources is that \citet{2022ApJ...930...39V} YSOs are on average 2.7 mag fainter in Gaia G band for any given distance and stellar mass bin. This indicates that these YSOs are more embedded, which could be interpreted as a sign of youth. However, we note there are other explanations as to why the historically known massive YSOs considered by us might be brighter (e.g. different evolution, envelope shredding, face-on disks, or pole-on sources exposing an outflow cavity).  An argument supporting a younger nature of the \citet{2022ApJ...930...39V} sample is that these sources typically have larger IR excess and H\textalpha{} emission by construction (see the discussion of \citealp{2020A&A...638A..21V}), which is classically understood as indicative of younger objects. 


Hence, an interpretation for the different behaviour of the two sets of massive YSOs ($>10$ M$_{\odot}$) in Fig.~\ref{Plot: Mosaic 2} is an evolutionary trend. The younger and more embedded massive YSOs are mostly clustered, whereas the older and optically brighter population is more isolated. However, we find it unlikely that in general massive YSOs would change their clustering properties so drastically in such a short timescale ($\lesssim 0.1$ Myr, \citealp{2012MNRAS.427..127B}). We believe some strong bias must also be at play. A possible scenario is that the historically known massive YSOs considered here, mostly first identified in old optical spectroscopic surveys (c.f. \citealp{1994A&AS..104..315T}), are heavily biased towards bright, little obscured, and isolated objects. Thus, they might not be representative of the true population of massive YSOs. This idea is reinforced by the fact that most massive YSOs are optically faint, due to the extinction from their surrounding envelopes (\citealp{2013ApJS..208...11L}). In contrast, \citet{2022ApJ...930...39V} sources were selected from \citet{2020A&A...638A..21V} catalog, which increased the number of known intermediate and high-mass YSOs with Gaia data by an order of magnitude, homogeneously identifying sources photometrically in an all-sky unbiased fashion with faint limit G$\sim21$ mag. If we assume the \citet{2022ApJ...930...39V} sources are more representative of the true population of massive YSOs, then the results of this work (Fig.~\ref{Plot: Mosaic 2}) indicate that the majority of the more massive YSOs are clustered. 

There is another evolutionary effect that deserves consideration. In Sect.~\ref{S_Vioque_Daniela} we compared the clustering rates of the sample of YSOs considered in this work with the clustering rates obtained via the same methodology for the sample of \citet[Fig.~\ref{Plot: Iglesias}]{2023MNRAS.519.3958I}. In this case, the comparison is limited to 7 M$_{\odot}$ and 400 pc. We see a clear difference in clustering rate between both samples. The typical values for our sample of YSOs are $\sim60\%$ and for \citet{2023MNRAS.519.3958I} are $\sim25\%$. \citet{2023MNRAS.519.3958I} sources have on average lower IR excesses and emission line strengths than our sample, which implies smaller disk radii and masses and smaller accretion rates. Hence, we expect their sources to be typically older than the ones of this work. This would imply we are tracing a decrease of the YSO clustering rate with time. There are, however, alternative explanations. The disk radius in intermediate-mass YSOs can be more dependent on the presence of giant planets than on time (see \citealp{2022A&A...662L...8P,2022AJ....164..109R,2022A&A...658A.112S, 2023A&A...669A.158S,2023A&A...671A.140G}), and although it is likely that the accretion rate decreases with time, episodic accretion seems to be the norm (\citealp{2022arXiv220311257F,2023arXiv230811430G}).

In conclusion, our results suggest that a large fraction of the more massive YSOs ($>10$ M$_{\odot}$) are clustered. In addition, we have evidence to suspect that intermediate and high-mass YSOs become less clustered with time, and –at least in the 1.5-5 M$_{\odot}$ range– in a fashion that does not vary much with stellar mass. This evolution towards isolation is probably dominated by cluster expansion and dispersion (\citealp{2009ApJS..184...18G,2019ApJ...870...32K, 2019ARA&A..57..227K, 2020SSRv..216...64K}). Cluster dissipation and disruption is likely caused by Galactic evolution (\citealp{2011MNRAS.414.1339K, 2018MNRAS.477.1683M}). However, we note that star cluster survivability after gas expulsion is independent of the impact of the Galactic tidal field (\citealp{2019MNRAS.486.1045S}). Hence, we expect the increase of isolated intermediate and high-mass YSOs with time to be independent of Galactic location (although see \citealp{2023MNRAS.524..555A}).






\subsection{Cluster properties of YSOs of different stellar masses}\label{S_massive}

For those intermediate and high-mass YSOs we detect clustered, we do not see any major correlation between cluster properties and the mass of the considered YSOs. There is also no correlation with the isochronal ages of the YSOs. The only result our analysis suggests is that YSOs above 4 M$_{\odot}$ are mostly in clusters larger than $2.5$ pc which are typically of low stellar density ($<3$ stars/pc\textsuperscript{2}, Fig.~\ref{Plot: Cluster_analysis}), but there are exceptions. 

In addition, Fig.~\ref{Plot: Cluster_analysis_2} shows that there is no major correlation between the mass of the YSOs and their projected distance to the geometrical centre of the clusters (in \textit{ra} \& \textit{dec}). We can only note that above 6 M$_{\odot}$ none of the sources are at very large distances ($>10$ pc) and there are hints of a decreasing trend of distance to cluster centre with YSO stellar mass. Above 12 M$_{\odot}$, all 6 clustered YSOs are within 3 pc of the cluster centre. Comparing to the cluster radius, most intermediate and high-mass YSOs are within $R_{80\%}$ (89\%) and $\sim15\%$ of them are in the inner 20\% region of the clusters. Again, no correlation with stellar mass is observed.

Following on from the conclusions of Sect.~\ref{Key_section}, we would expect in a competitive accretion like scenario the more massive YSOs to appear in large clusters, and closer to the centre of the clusters. Our results might indicate that the very massive YSOs ($>10$ M$_{\odot}$) are not in small clusters (c.f. \citealp{1999MNRAS.309..461B}), and that they are closer to the centre of the clusters. However, more data points are needed to confirm this trend (see Sect.~\ref{S_dis_more_massive} for the caveats of the high-mass regime). Competitive accretion also predicts that the mass of the most massive star in a cluster increases as the system grows in number of stars and in total mass (\citealp{2004MNRAS.349..735B}). However, we see no correlation between the amount of stars per cluster and the mass of the considered YSO (Fig.~\ref{Plot: Cluster_analysis}). In contrast, a monolithic collapse like scenario predicts strongly peaked density distributions. We can test this in 2d sky space by using the tidal radius (\citealp{1962AJ.....67..471K}), or the radius at which the cluster has the best contrast to field stars. If a monolithic collapse like formation dominates at high-masses, we would expect to see that low mass YSOs are typically in clusters with larger tidal radii. However, we do not report any correlation between the tidal radii and the mass of the considered YSOs.

Therefore, the analysis of the properties of the clusters identified in this work is not conclusive of a competitive accretion like scenario acting in the most massive YSOs ($>10$ M$_{\odot}$). Likewise, we find no argument from this analysis supporting a more isolated, monolithic-type formation for the more massive YSOs. An explanation for these results could be that mass segregation happens at later stages of cluster lifetimes (\citealp{2023A&A...674A..93D,2023MNRAS.523.2083F}), and that cluster properties at early stages are determined by the initial cloud size and mass, and not the stellar content (\citealp{2023MNRAS.523.2083F,2023ApJ...950..148M}). It is also an option that the YSOs considered in this work are not the most massive YSOs in several of the detected clusters, which might contain heavily embedded massive YSOs that escaped the Gaia survey.


\subsection{Comparison with previous results}


Our results support the main conclusion of the seminal works of \citet{1995PhDT.........1H} and \citet{1997A&A...320..159T, 1999A&A...342..515T}. There is tentative evidence to suspect that YSOs above $10$ M$_{\odot}$ appear more clustered than intermediate-mass YSOs. They based their results in a 2d near-IR K band clustering analysis. This work was later extended in \citet{2003A&A...400..575H}, \citet{2008ApJ...685.1005Q}, and \citet{2009A&A...503..801F}. We now analyse the clustering properties of intermediate and high-mass YSOs in the Gaia era, providing an analysis in 3d and 5d. To our knowledge, we are the first to characterize the transition between the clustering properties of low-mass and intermediate-mass YSOs (the $1.5-10$ M$_{\odot}$ range). Our findings in the $4-10$ M$_{\odot}$ regime also support the idea that the formation of high-mass stars is influenced by dynamical interaction in a young cluster environment. However, we suggest that the transition in clustering properties at $\sim10$ M$_{\odot}$ is not smooth, contrary to what was reported in \citet{1999A&A...342..515T}. In addition, we show that YSOs above $10$ M$_{\odot}$ have similar or higher clustering rates than T Tauri stars, being the difference easily explained by the shorter timescale of massive YSO evolution. Contrary to \citet{1999A&A...342..515T}, we see no transition from low density aggregates of T Tauri stars to dense clusters around massive stars. 

Something to consider are the different scales analysed by different studies. \citet{1995PhDT.........1H}, \citet{1999A&A...342..515T}, \citet{2003A&A...400..575H}, \citet{2008ApJ...685.1005Q}, and \citet{2009A&A...503..801F} often report clusters with radii of 0.2 to 1 pc, whereas most of the clusters identified in this work have typical values of $2-15$ pc (Fig.~\ref{Plot: Cluster_size}). Another significant difference is the number of stars per cluster. We have identified clusters with a few tens to a few hundred members, whereas previous works mostly report clusters with an order of magnitude fewer members.  We believe these differences are mainly caused by the difference in methodologies. Previous authors mostly sought clusters in the nearby environments of the intermediate and high-mass YSOs. This has proven useful in many science cases (e.g. \citealp{2020MNRAS.494.5851S,2021MNRAS.507..267A}), but this methodology favours peak-density results in the vicinity of the considered YSOs. In contrast, we have applied a position-independent methodology to the whole Gaia space, without favouring any specific mass range (Sect.~\ref{S_HDBSCAN}). \citet{2008MNRAS.391..711M} already alerted that previously reported clusters were too sparsely populated to match random drawing from the IMF. Our results fit the expected theoretical values of \citet{2008MNRAS.391..711M}, and the clusters we have obtained have similar properties to the Galactic population of open clusters.  One more significant difference with previous works is that our analysis is optically based, whereas previous works are mostly near-IR based. This does not ease a direct comparison, as in Gaia we are missing sources fainter than $G\sim21$ mag (see Appendix~\ref{Appendix_B}). We conclude that in this work we have retrieved the complete scale of the clusters containing intermediate and high-mass YSOs, whereas previous works have mostly provided a more complete (in number) yet limited (in volume) view of the vicinity of these YSOs. Both analyses are hence complementary and necessary to achieve high-purity studies like that of \citet{2011ApJ...727...64K} for low-mass stars in the vicinity of the Sun.


%

In this work, we report 18 isolated YSOs with $M>10$ M$_{\odot}$ in Gaia astrometric space (four of them from \citealp{2022ApJ...930...39V}). Regarding the possibility of truly isolated massive star formation (i.e. in the absence of a cluster environment, we note that $\sim50-70\%$ of intermediate-mass YSOs are binaries, \citealp{2006MNRAS.367..737B,2023AJ....165..135T}), several candidates have been put forward (e.g. \citealp{2022ApJ...939..120L, 2023ApJ...942....7F}), although in depth analyses have found small clusters around most massive YSOs seemingly isolated (\citealp{2017ApJ...834...94S, 2021MNRAS.507..267A, 2023A&A...670A.151Y}). Similarly, although isolated O stars have been reported (e.g. \citealp{2005A&A...437..247D,2013ApJ...768...66O}), this is disputed and it often seems to be a result of observational limitations (\citealp{2007MNRAS.380.1271P,2017ApJ...834...94S}). Theoretical analysis show that seemingly isolated O stars might be explained by low-mass clusters sampled randomly from a standard initial mass function (\citealp{2007MNRAS.380.1271P}). 






\section{Conclusions}\label{S_Conclusion}

This study constitutes the first large scale analysis of the clustering properties of intermediate and high-mass YSOs ($1.5-20$ M$_{\odot}$), encompassing the classical groups of Herbig Ae/Be stars, Massive Young Stellar Objects (MYSOs), and Intermediate-Mass T Tauris. We applied an HDBSCAN clustering methodology to 337 spectroscopically characterized intermediate and high-mass YSOs with Gaia data. We analysed the clustering properties of the 263 stars located in the distance range where our methodology works best ($350-4000$ pc). We present the resulting cluster parameters for these sources in Table~\ref{Table}. The main conclusions we report are:

\begin{itemize}
    \item Intermediate-mass YSOs in the range $4-10$ M$_{\odot}$ are more isolated than T Tauri stars and Herbig Ae/late-Be stars ($<4$ M$_{\odot}$) and MYSOs ($>10$ M$_{\odot}$). The clustering rate in this intermediate-mass regime decreases as a function of stellar mass. We propose that this is due to the fact that most YSOs above $10$ M$_{\odot}$ demand the presence of clusters of stars around them, which contain mostly low-mass stars, and hence both groups have high clustering rates. In contrast, YSOs in the $4-10$ M$_{\odot}$ regime are too massive to appear often in these MYSO-induced clusters, but not massive enough to have their own clusters. We propose they tend to form similarly to T Tauri stars, via a monolithic collapse-like scenario. Hence, intermediate-mass star formation in the range $4-10$ M$_{\odot}$ is more isolated than star formation at other mass regimes.
    
    \item Of the 263 stars, 71 are clustered in 3d (27\%, position and parallax), 92 are clustered in 5d (35\%, position, parallax, and proper motion), 36 are clustered in both 3d and 5d (14\%), and 127 are clustered in 3d or 5d (48\%). A total of 136 intermediate and high-mass YSOs are isolated (52\%), not appearing clustered in any Gaia astrometric space. These clustering rates vary with stellar mass. In particular, we report a $\sim55-60\%$ clustering rate for the lower mass YSOs ($1.5-4$ M$_{\odot}$), and a clustering rate as low as $\sim27\%$ for intermediate mass YSOs  in the range $4-10$ M$_{\odot}$. These percentages are likely lower-limits of the true clustering rates due to the optical limitation of the Gaia mission. However, we show that the conclusions of this work are not affected by this limitation.

    
    \item We observe that for YSOs $>10$ M$_{\odot}$ there is a significant difference in clustering properties between the YSOs homogeneously identified in \citet{2022ApJ...930...39V} and those historically considered. \citet{2022ApJ...930...39V} MYSOs have a high clustering rate, whereas the historically known sources are mostly isolated. \citet{2022ApJ...930...39V} sources show signs of a younger population. Hence, we believe this difference is due to a mixture of an evolutionary effect and a historical bias. If we assume that the \citet{2022ApJ...930...39V} sample is more representative of the population of massive YSOs in the Galaxy, then we can conclude that most massive YSOs appear clustered, and that this clustered nature starts at around $\sim10$ M$_{\odot}$. We report 18 isolated YSOs with  $M>10$ M$_{\odot}$ in Gaia astrometric space (4 of them from \citealp{2022ApJ...930...39V}). We warn the reader that above $13$ M$_{\odot}$ low-number statistics and unknown population biases might be affecting the reported clustering rates.
    
    \item We find that intermediate and high-mass YSOs become less clustered with decreasing disk emission and accretion rate. This points towards an evolution with time. In the $1.5-5$ M$_{\odot}$ range this evolution with time towards isolation does not vary much with stellar mass.
    
    \item We find no major correlation between the stellar properties of the YSOs considered and the properties of their clusters. We find no conclusive evidence from the cluster properties that can support either competitive accretion or monolithic collapse theories at high stellar masses. We can only report a weak trend of more massive stars being closer to cluster centres above $6$ M$_{\odot}$, and that YSOs above $4$ M$_{\odot}$ are often in clusters larger than 2.5 pc in radius. These clusters tend to have medium to low stellar densities ($<3$ stars/pc\textsuperscript{2}, in Gaia space).
    
    \item We present 55 new clusters not reported before in the literature. We note that most of these new clusters are larger and less populated than what is normal in open clusters. Hence, they often do not fit the classical definition of open cluster, being more similar to YSO sparse associations. The other clusters detected in this work have been previously identified with similar sizes and number of members. Eleven clusters contain more than one intermediate or high-mass YSO.

\end{itemize}

The results of this work are based on almost the complete population of intermediate and high-mass YSOs which currently have spectroscopic characterization (337 sources). As it is discussed in the text, this sample might be affected by some selection biases, and it is limited in the high-mass range ($>13$ M$_{\odot}$). Many of the intermediate and high-mass YSO candidates of \citet{2020A&A...638A..21V} catalogue will be observed by the WEAVE spectrograph (\citealp{2023MNRAS.tmp..715J}). This will provide the community with an order of magnitude more sources to conduct studies like the one presented in this work (\citealp{2022ApJ...930...39V} estimated a catalogue accuracy of $>90\%$). This will provide key statistical insights to some of the discussion presented here.

\begin{acknowledgments}
\section*{Acknowledgments}
We thank Uma Gorti, Cathie Clarke, and Jesús Maíz Apellániz for useful discussions that contributed to this work. Álvaro Ribas has been supported by the UK Science and Technology research Council (STFC) via the consolidated grant ST/W000997/1 and by the European Union’s Horizon 2020 research and innovation programme under the Marie Sklodowska-Curie grant agreement No. 823823 (RISE DUSTBUSTERS project). Ignacio Mendigutía's research is supported by a “Ramón y Cajal” grant (RyC2019-026492-I), funded by MCIN/AEI/10.13039/501100011033 and by the “FSE invierte en tu futuro”. This work presents results from the European Space Agency (ESA) space mission Gaia. Gaia data are being processed by the Gaia Data Processing and Analysis Consortium (DPAC). Funding for the DPAC is provided by national institutions, in particular the institutions participating in the Gaia MultiLateral Agreement (MLA). This research has made use of the TOPCAT tool (\citealp{2005ASPC..347...29T}). The Geryon
cluster at the Centro de Astro-Ingenieria UC was extensively used for the 
calculations performed in this paper. BASAL CATA PFB-06, the Anillo ACT-86, 
FONDEQUIP AIC-57, and QUIMAL 130008 provided funding for several improvements 
to the Geryon cluster.
\end{acknowledgments}

\appendix

\section{Biases and incompleteness}\label{Appendix_B}

In this appendix, we discuss in detail the possible sources of biases and incompleteness present in this work.

\subsection{Assessment of Gaia 5d space incompleteness}\label{S_IR_images}

We have based our analysis in the sources with a 5-parameter astrometric solution in Gaia DR3. Here, we evaluate how many sources are we missing by not selecting all the sources present in Gaia (i.e., with 2-d solutions, \textit{ra} and \textit{dec}). We also consider the possible impact of the Gaia astrometric uncertainties.

\begin{figure}[t!]
\centering\includegraphics[width=\columnwidth]{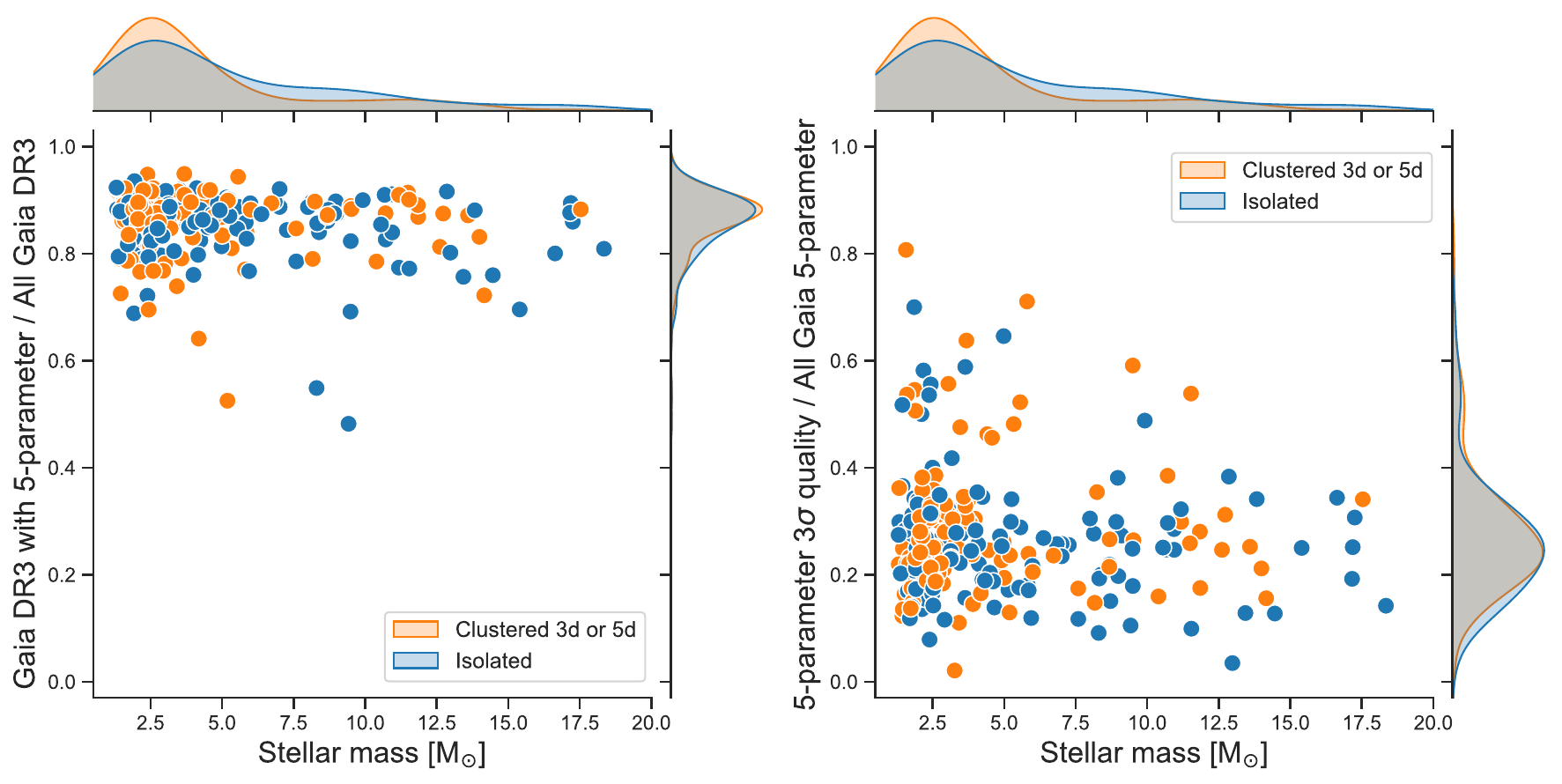}
\caption{Within 3 arcminute of each intermediate or high-mass YSO. \textit{Left:} Fraction of sources considered in our analysis (with Gaia DR3 5-parameter solution) with respect to all Gaia DR3 sources. \textit{Right:} Fraction of Gaia DR3 sources with all 5 astrometric quantities having values over three times their uncertainty with respect to the sources considered in our analysis. The stellar mass of the intermediate or high-mass YSOs is shown in the horizontal axes. Density curves of clustered and isolated sources are shown at the top and right.}\label{Plot: Gaia_5_all}
\end{figure}

To have a meaningful estimation of the sources missed around every intermediate and high-mass YSO, we limit the radius of comparison to 3 arcminutes. This angular radius roughly corresponds to half $R_{80\%}$ radius for most clusters considering typical distances of $1000-2000$ pc (Fig.~\ref{Plot: The sample distance}). The results of this study are shown in the left panel of Fig.~\ref{Plot: Gaia_5_all}. On average, we have missed 12\% of the Gaia sources by demanding a 5-parameter astrometric solution. However, there is no dependence between YSO clustered nature or stellar mass and the amount of Gaia sources missed by demanding a 5-parameter astrometric solution. We conclude that demanding a 5-parameter solution does not affect our classification of clustered or isolated YSO nature, and neither it affects the trends with stellar mass detected in Sect.~\ref{S_Clus_vs_mass}.

The Gaia astrometry of the considered sample is of high-quality, i.e. there are no intermediate or high-mass YSOs in our sample with large astrometric uncertainties (Sect.~\ref{S_sample}). This is true for relative errors ($>3\sigma$), and for absolute uncertainty values (e.g., proper motion errors are under 0.5 mas/yr in all cases). In addition, we report no correlation in our sample of YSOs between stellar mass and astrometric uncertainties. In the right panel of Fig.~\ref{Plot: Gaia_5_all}, we evaluate how many neighbouring sources to each intermediate and high-mass YSO have astrometric values larger than three times their uncertainty (in ra, dec, parallax, proper motion ra, and proper motion dec). While we find that, in general, only $\sim30\%$ of the sources used to find clusters with HDSBCAN have more than  3\textsigma{} astrometric quality, again we report no significant correlation between the astrometric quality of the sources used for HDBSCAN and the intermediate or high-mass YSOs stellar masses or derived clustered nature. Hence, we claim that if astrometric uncertainties are blurring structures together this effect is very small in our derivations, and in any case affecting equally all the stellar mass ranges considered.


Gaia has another important limitation for clustering studies. This is that it is limited to optically bright sources ($G<21$ mag), and it does not contain the more embedded IR-bright YSOs. Hence, some of the seemingly isolated intermediate and high-mass YSOs of our sample might be surrounded by many low-mass stars that Gaia cannot detect. Indeed, cluster membership analysis with the near-IR VVV survey (VISTA Variables in the Via Lactea, \citealp{2010NewA...15..433M, 2012A&A...537A.107S}) have found $\sim45\%$ more member candidates in open clusters than those found only using Gaia (\citealp{2021MNRAS.503.1864P,2022MNRAS.513.5799P}).


\begin{figure*}[t!]
\includegraphics[width=\textwidth]{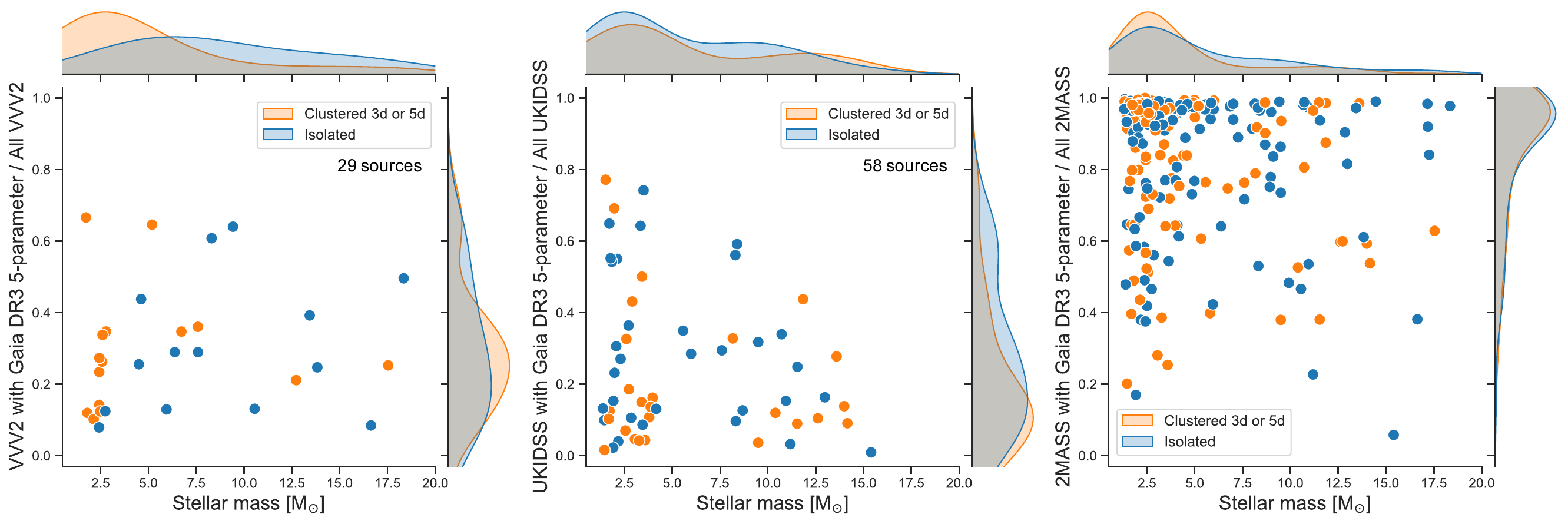}
\caption{Within 3 arcminute of each YSO, this plot shows the fraction of sources detected by IR surveys which were considered in our analysis (those with Gaia DR3 5-parameter solution). From left to right, VVV, UKIDSS, and 2MASS. Horizontal axes show the stellar mass of each intermediate or high-mass YSO. Density curves of clustered and isolated sources are shown at the top and right.}\label{Plot: IR_surveys}
\end{figure*} 

In order to assess the impact of this limitation in our analysis, we use the VVV Data release 2. There are 29 sources of our sample in the footprint of VVV, roughly $\sim11\%$ of our sample. As before, we made a 3 arcminute search around each of our YSOs, and studied the fraction of sources detected by VVV which also have a Gaia 5-parameter solution. The result is shown in the left panel of Fig.~\ref{Plot: IR_surveys}. In the same Figure we present an equivalent analysis for the UKIDSS-DR6 Galactic Plane Survey (\citealp{2008MNRAS.391..136L}, 58 sources match with our sample) and 2MASS (\citealp{2006AJ....131.1163S}, all the sources of our sample are in 2MASS). These three surveys span from J band ($1.2\;\mu m$), to K\textsubscript{s} band ($2.2\;\mu m$), with different optical depths. In the 2MASS comparison we obtain that, on average, 95\% of the sources around our YSOs are also in Gaia 5-parameter space. However, in the comparison with VVV and UKIDSS we see that, on average, only 20-30\% of the sources around each YSO have a Gaia 5d counterpart. In any case, we see in Fig.~\ref{Plot: IR_surveys} that for the VVV, UKIDSS, and 2MASS comparisons there is no significant correlation between Gaia 5d incompleteness and the clustered and isolated populations, and neither there is a correlation with the stellar mass of the intermediate or high-mass YSOs. In addition to this, the depth and wavelength coverage of these surveys imply that most of the sources Gaia is not detecting are likely unrelated to the intermediate or high-mass YSOs of interest. It is hard to assess what fraction of those sources that we are missing are real companions of the considered YSOs, as these IR surveys go very deep (e.g., VVV faint limit is K\textsubscript{s}=17 mag) and we lack parallax information for their sources. We note that these IR surveys are limited to the Galactic centre, where the extinction is largest, and hence they suppose an upper-limit to the Gaia incompleteness.




\begin{figure*}[t!]
\includegraphics[width=\textwidth]{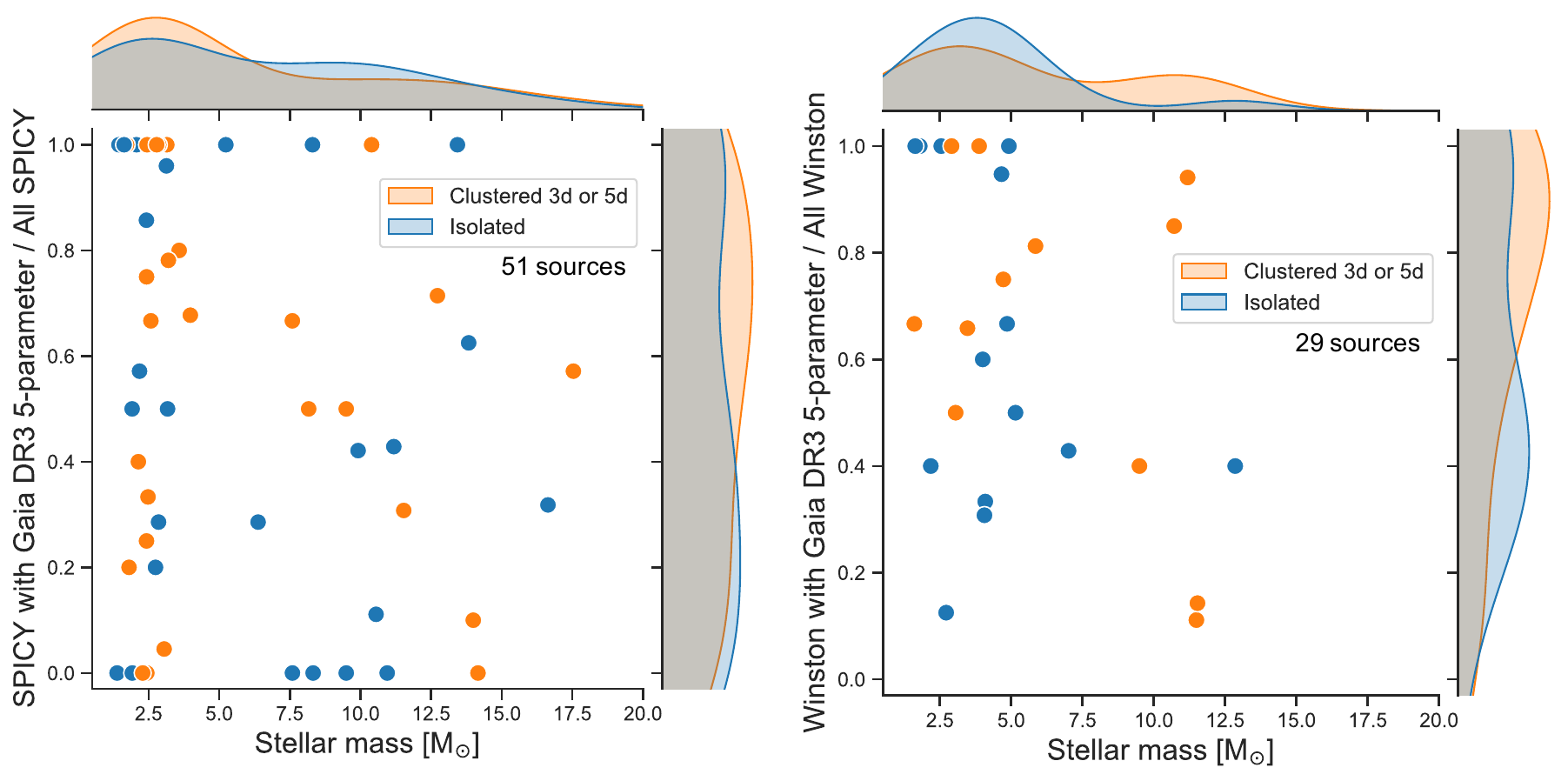}
\caption{Within 3 arcminute of each YSO, this plot shows the fraction of sources appearing in the IR-based catalogues of \citet[SPICY]{2021ApJS..254...33K} and \citet{2020AJ....160...68W} which were considered in our analysis (with Gaia DR3 5-parameter solution). Horizontal axes show the stellar mass of each intermediate or high-mass YSO. Density curves of clustered and isolated sources are shown at the top and right.}\label{Plot: YSO_surveys}
\end{figure*} 




%


Another way to assess the limitations of the Gaia 5d space is to compare it with infrared-based YSO catalogues. For this, we use the SPICY (\citealp{2021ApJS..254...33K}) and \citet{2020AJ....160...68W} catalogues. SPICY catalogue contains 117446 YSOs in the inner Galactic midplane. This catalogue was produced using mainly Spitzer data, but also 2MASS, VVV and UKIDSS data, being the most comprehensive catalogue of YSOs from  mid-infrared observations. A similar catalogue was produced by \citet[43094 sources]{2020AJ....160...68W}, in this case targetting the outer Galaxy, making these two catalogues complementary. The accuracy of these catalogues has been evidenced by spectroscopic surveys (e.g. \citealp{2023AJ....165....3K}). Of our sample of 263 sources, 51 sources overlap with SPICY and 29 with \citet{2020AJ....160...68W} catalogue. We study what is the fraction of sources appearing in these IR-based catalogues which were considered in our analysis. Given that these catalogues have a limited footprint, and for consistency with the previous analyses, we also consider a 3 arcminute aperture to look for missing sources in the Gaia 5-parameter space. The results of this study are shown in Fig.~\ref{Plot: YSO_surveys}. In this comparison, we also get clustered stars at high and low levels of Gaia 5d completeness, and the means of the distributions of clustered and isolated stars are within 1-sigma of each other. In addition, no bias with respect to the stellar mass of the intermediate or high-mass YSOs can be appreciated. We note that these IR surveys might be showing unrelated background or foreground YSOs. Indeed, only 67\% of the SPICY sources with Gaia parallaxes are on average within 1000 pc of the corresponding intermediate or high-mass YSO. 



Our results in this appendix give us no reason to believe we would have different incompleteness levels around sources of different masses. We conclude that the evolution of the clustering ratio as a function of stellar mass presented in Sect.~\ref{S_Clus_vs_mass} is robust (Fig.~\ref{Plot: Mosaic 1} and~\ref{Plot: Mosaic 2}), and it is not affected by Gaia 5d incompleteness or astrometric uncertainty. We, however, cannot claim the absolute clustering rate values presented in Sects.~\ref{S_HDBSCAN} and~\ref{S_Analysis}, and Figs.~\ref{Plot: Mosaic 1},~\ref{Plot: Mosaic 2}, and~\ref{Plot: Iglesias} to be accurate, as they are lower limits, given that we are missing an unknown fraction of sources around each YSO because of not being present in Gaia 5-parameter astrometric space.

\subsection{Assessment of possible distance and sky position biases}\label{Distance_effects}

\begin{figure}[t!]
\centering\includegraphics[width=\textwidth]{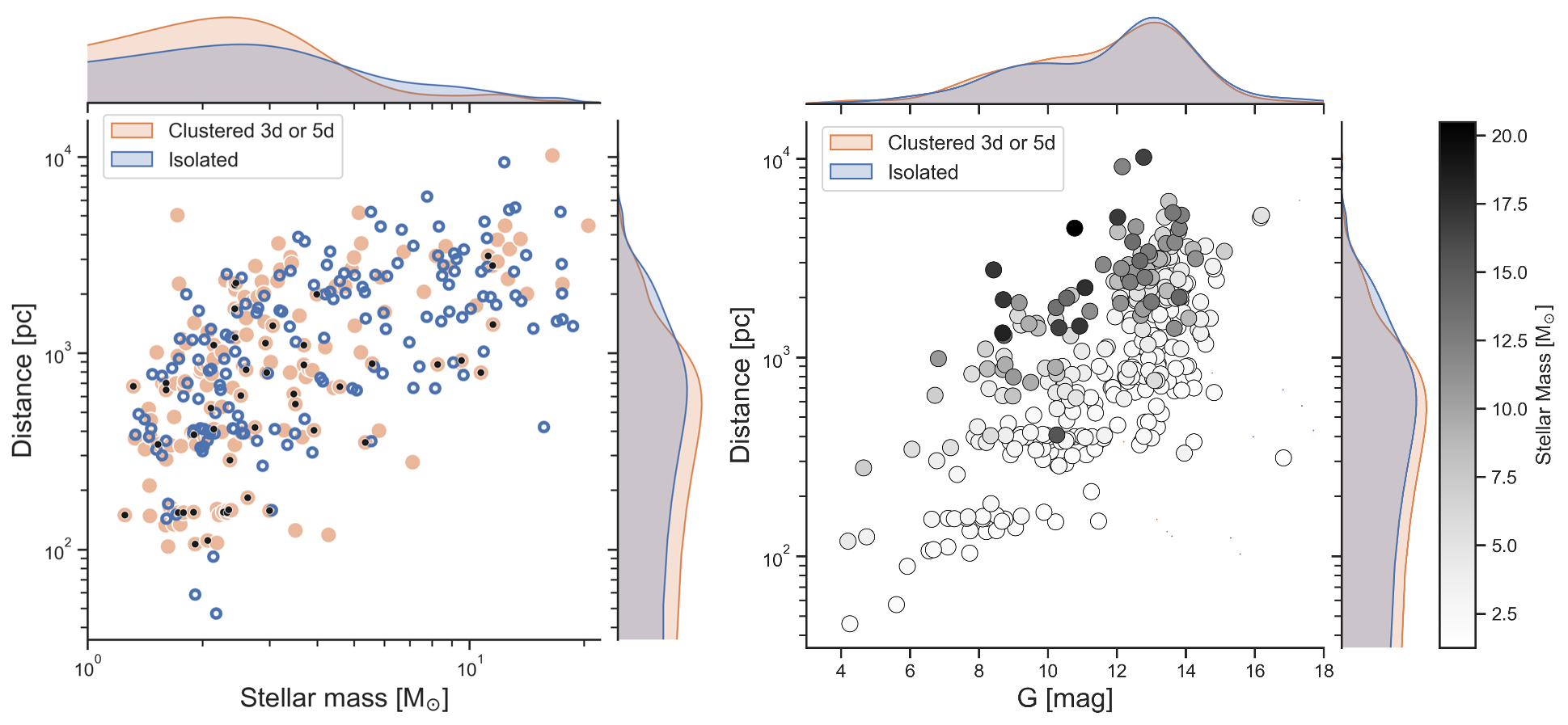}
\caption{\textit{Left:} Distance vs. stellar mass for all YSOs compiled in this work (337 sources). Density curves of clustered and isolated sources are shown at the top and right. Black dots indicate those sources that appear clustered both in 3d and 5d. \textit{Right:} Distance vs. Gaia G band magnitude for all YSOs  compiled in this work. Colour scheme indicates stellar mass, with darker sources being more massive (linear scale). For the analysis we have only considered sources between 350 and 4000 pc.}\label{Plot: Distance_effects_1}
\end{figure}

\begin{figure*}[t!]
\includegraphics[width=\textwidth]{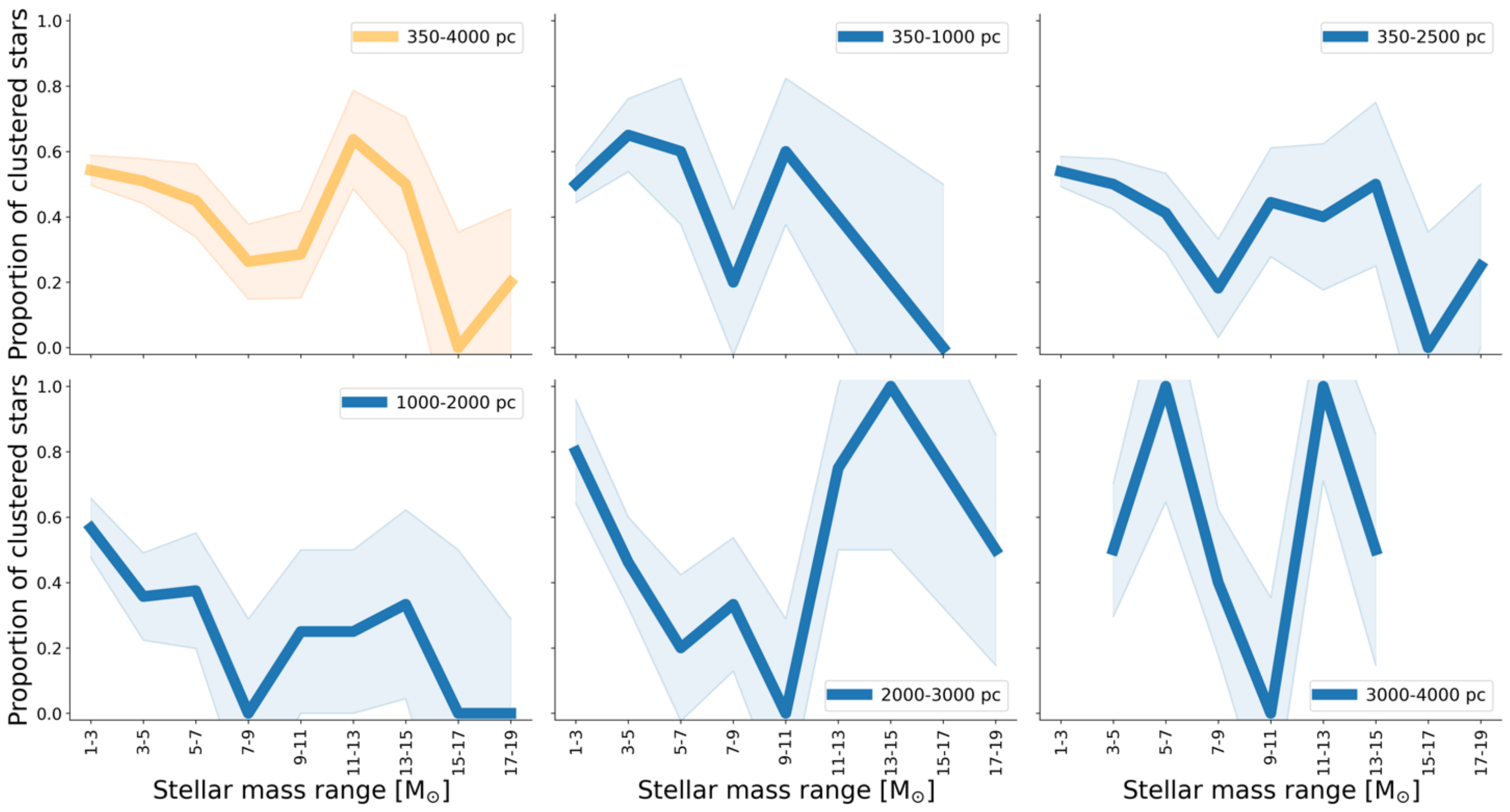}
\caption{Same as Fig.~\ref{Plot: Mosaic 1} `3d or 5d' with 2 M$_{\odot}$ bins but considering only certain distance ranges. Top left panel is identical to Fig.~\ref{Plot: Mosaic 1} bottom left `3d or 5d' panel. We show it here for comparison purposes. Complete error is $1/\sqrt{n}$, with $n$ being the number of sources per bin.}\label{Plot: Distance_effects_2}
\end{figure*}

According to \citet{2021A&A...646A.104H} and our Sect.~\ref{Sect. HDBSCAn results}, our HDBSCAN methodology should be efficient and consistent in the distance range considered in this work ($350-4000$ pc), and it should be independent of sky location. In this section, we evaluate possible distance and sky location effects in our clustering derivations.

We start by noting that our YSO sample is well distributed in stellar mass across the whole distance range considered (Fig.~\ref{Plot: Distance_effects_1}). In the distance range mainly analysed in this work ($350-4000$ pc), we have both clustered and isolated stars across all distances. This is also true for stars that appear clustered in both 3d and 5d. In addition, there is no significant difference between clustered and isolated stars as a function of distance (Fig.~\ref{Plot: Distance_effects_1}). We report no distance effect on the measured cluster properties: number of stars per cluster ($N_{c}$),  cluster radii ($R_{80\%}$), and cluster densities. In addition, we detect clustered stars at all proper motion ranges. Hence, the methodology is not biased against high proper motion stars.

In the distance range mainly analysed in this work ($350-4000$ pc), there are two noticeable effects. First, we lack massive YSOs (above 8 M$_{\odot}$) below 500 pc. This is a real, purely IMF effect, massive stars form fast and are rare, so there are very few of them nearby with Gaia data. The other effect is artificial, we lack low-mass YSOs (1.5-3 M$_{\odot}$) above 2500 pc. This is due to the fact that these sources are too faint to have fallen in the spectroscopic surveys used to construct the sample.  In order to assess how much the two previous distance effects affect our results, we repeat the analysis of Sect.~\ref{S_Clus_vs_mass} for different distance ranges ($350-1000$, $350-2500$, $1000-2000$, $2000-3000$, and $3000-4000$ pc, Fig.~\ref{Plot: Distance_effects_2}). The general shape of the evolution of clustering rate with stellar mass described in Sect.~\ref{S_Clus_vs_mass} holds independently of the distance range considered. It goes down from lower-mass sources to the intermediate-mass regime, and then goes up again for the more massive YSOs. The mass bin of minimum clustering rate varies, but it is always contained within the $7-10$ M$_{\odot}$ range. At the more massive end ($M>13-15$ M$_{\odot}$), low number statistics and population biases affect the results, and the clustering rates presented are dominated by very few sources (see Sect.~\ref{S_dis_more_massive} for further discussion). Outside the distance range mainly analysed in this work ($350-4000$ pc), Fig.~\ref{Plot: Distance_effects_1} shows that there is a strong bias favouring cluster detection for distances under 200 pc.

There is one more distance effect we have detected for the lower-mass YSOs. Beyond 1500 pc most low-mass stars belong to the \citet{2022ApJ...930...39V} sample, which by construction has a Gaia G brightness faint limit of $\sim14.5$ mag. Therefore, at larger distances the lower-mass objects we are including in the analysis are forced to be optically more luminous than those at short distances. This implies that they are on average less extincted. Our methodology finds it easier to detect clusters around YSOs in low extinction environments, as more sources will appear in Gaia 5d space (Sect.~\ref{S_IR_images}). We hence expect to detect a higher fraction of clustered low-mass YSOs at distances above $1500$ pc. This is indeed what we see in Fig.~\ref{Plot: Distance_effects_2}. Lower-mass YSOs have higher clustering rates at $2000-4000$ pc distances (of $\sim80\%$) than at lower distances. We note that we report a clustering rate of $\sim55-60\%$ for the lower mass stars until 2000 pc, which matches the one reported for the complete sample. Hence, this bias affects a small fraction of the lower-mass sources and does not have an impact on our conclusions.


No correlation between clustering properties or YSO stellar mass and sky coordinates or Galactic latitude or longitude was found. Hence, we do not expect the Gaia selection function to affect our results (\citealp{2023A&A...669A..55C}). The only sky related difference in the methodology appeared in Sect.~\ref{S_s_HDBSCAN}, where we decided to use HEALPix 6 instead of HEALPix 5 for sources which were in very crowded fields, to retrieve similarly sized clusters as for the rest of sources. Of the sample of 263 sources considered in Sect.~\ref{Sect. HDBSCAn results}, we have 191 stars with HEALPix 5 and 72 with HEALPix 6.  85\% of these 72 HEALPix 6 stars are beyond 1 kpc and hence, in general, we can safely assume the HEALPix choice does not make a difference as to whether a given source was identified as part of a cluster or not. In any case, we have made a comparison between both HEALPix subsets. These sets have 49\% (HEALPix 5) and 47\% (HEALPix 6) of clustered stars. Hence, HDBSCAN is equally efficient in both cases. As the more massive sources are typically further away and hence more likely situated against the crowded Galactic plane, we have also evaluated whether this efficiency varies as a function of stellar mass. We observe that this is not the case. The cluster properties obtained with HEALPix 5 and HEALPix 6 are identical.


We thus conclude that the distance and sky distribution of the considered sample of intermediate and high-mass YSOs do not affect the analysis and conclusions of this work.

\section{Assessment of other HDBSCAN parameters}\label{Appendix_A}

For a subsample of 235 sources, we have applied HDBSCAN to the 5 dimensional astrometric space using a wider grid of hyperparameters (`min\textunderscore cluster\textunderscore size' and `min\textunderscore samples'). We used a grid defined as [5,5], [5,10], [10,5], [10,10] (or [5-10, 5-10] step 5) and [60-100, 60-100] (step 5). We ran HDBSCAN at each field following the grid order, stopping if the cluster probability for the massive YSO was different to $0$. This allowed us to assess the limitations of Section~\ref{S_s_HDBSCAN} methodology, considering smaller and bigger cluster sizes as well as variations of `min\textunderscore samples'.


For 197 of the 235 sources ($84\%$) there is agreement between the grids of Section~\ref{S_s_HDBSCAN} and the one of this section, either if the stars are clustered or unclustered. There are only three stars ($1\%$) that were identified with the [30, 30] setting that were not identified by this section's grid. However, there are 35 stars ($15\%$) that were identified as part of a cluster with this section's grid that were not detected in Section~\ref{S_s_HDBSCAN}. Of those 35, 16 appear clustered with a minimum cluster size of 5, and 10 required of a minimum cluster size of 100 to be detected as part of a cluster. These two extremes are likely not indicative of physical clusters. A minimum cluster size of 5 as traced by HDBSCAN cannot possibly be statistically robust (minimum cluster size is often considered to be 10, \citealp{2019ARA&A..57..227K, 2020SSRv..216...64K}). Similarly, if the clusters detected with minimum cluster size of 100 were real, they should have been detected at smaller cluster sizes (c.f. \citealp{2021A&A...646A.104H} which proves `min\textunderscore cluster\textunderscore size' from 10 to 80). Therefore, there are only 9 sources that can be truly considered as missed clusters in Section~\ref{S_s_HDBSCAN} (seven with [10, 5] and two with [60, 60]).


Hence, the choice of hyperparameters in Sect.~\ref{S_s_HDBSCAN} is not affecting the determination of whether a star is clustered or not to within $95\%$ confidence. Therefore, we conclude that the hyperparameters chosen in this work are appropriate for identifying clusters and associations with HDBSCAN around intermediate and high-mass YSOs.

\section{Table of clustered and isolated stars}

In this table, we detail the clustering properties of the 263 intermediate and high-mass YSOs analysed between 350 and 4000 pc. When it corresponds, the cluster properties are also tabulated. Here we present a portion of the table for guidance regarding its form and content. The full table will be made available online at the CDS in the VizieR archive service.


\begin{longrotatetable}
\begin{deluxetable*}{llllcccccccccl}
\tabletypesize{\scriptsize}
\tablecolumns{14}
\setlength{\tabcolsep}{4pt}
\label{Table}
\tablecaption{Clustering properties of the 263 intermediate and high-mass YSOs analysed between 350 and 4000 pc. When it corresponds, the cluster properties are also tabulated.}
\tablehead{
\colhead{SIMBAD} & \colhead{Gaia DR3} & \colhead{RA} & \colhead{DEC}  &  \colhead{Stellar mass} & \colhead{Distance} & \colhead{Gaia G} & \colhead{HDBSCAN} & \colhead{HDBSCAN} & \colhead{$R_{80\%}$} & \colhead{Members} & \colhead{Density} & \colhead{Distance} & \colhead{Notes}\\
\colhead{name} & \colhead{source id} & \colhead{hh:mm:ss} & \colhead{dd:mm:ss.s}  & \colhead{[M$_{\odot}$]} & [pc] & \colhead{[mag]} & \colhead{probability} & \colhead{configuration} & \colhead{[pc]} & & stars/$pc^2$ & \colhead{to centre [pc]} & \colhead{}}
\startdata
HD  36917 & 3017264007761349504 & 05:34:47 & -05:34:14.6 & $3.64^{0.58}_{0.29}$ & $447.95^{11.32}_{10.21}$ & 7.96 & 0 & Isolated & - & – & – & – & –  \\
HD 245185 & 3338005774514542720 & 05:35:10 & +10:01:51.4 & $2.14^{0.26}_{0.05}$ & $410.26^{4.94}_{5.71}$ & 9.86 & 0.75, 0.84 & 3d, 5d & 9.66 & 702 & 1.92 & 3.26 & – \\
HD  36982 & 3017360348171372672 & 05:35:10 & -05:27:53.2 & $5.80^{0.58}_{0.58}$ & $404.00^{3.58}_{4.06}$ & 8.34 & 0.68 & 3d & 0.97 & 706 & 191.69 & 0.53 & cont  \\
NV Ori & 3017346952181227776 & 05:35:31 & -05:33:08.9 & $1.90^{0.12}_{0.08}$ & $383.61^{2.87}_{3.06}$ & 9.7 & 0.49, 0.84 & 3d, 5d & 3.58 & 675 & 13.38 & 1.44 & –  \\
T Ori & 3017347673735214592 & 05:35:50 & -05:28:34.9 & $2.38^{0.16}_{0.10}$ & $398.95^{5.34}_{4.46}$ & 11.29 & 0 & Isolated & - & – & – & – & –  \\
BD-06  1253 & 3016923017420354688 & 05:36:25 & -06:42:57.7 & $2.50^{0.31}_{0.11}$ & $379.64^{14.12}_{16.00}$ & 10.53 & 0 & Isolated & - & – & – & – & –  \\
... & ... &  ... & ... & ... & ... & ... & ... & ... & ... & ... & ... & ... & ... \\
\enddata
\vspace{-1cm}
\end{deluxetable*}
Coordinates and G magnitudes were obtained from Gaia DR3. Distances to sources are the geometric distances of \citet{2021AJ....161..147B}. Rest of parameters were derived in this work. In the \textit{Notes} column, `cont' means possible cluster miss-identification and `new' a new cluster identification previously unreported in the literature (c.f. \citealp{2023A&A...673A.114H}). Here we present a portion of the table for guidance regarding its form and content. The full table will be made available online at the CDS in the VizieR archive service. In the online version the four HDBSCAN resulting cluster probabilities from Sect.~\ref{S_s_HDBSCAN} appear tabulated in different columns, and Gaia DR3 coordinates in decimal degrees are also provided.
\end{longrotatetable}

\bibliographystyle{aasjournal}



\end{document}